\documentclass[aps,pra,twocolumn,superscriptaddress]{revtex4-2}

\usepackage{amsfonts,mathrsfs,amsmath,amsthm,amssymb}
\usepackage{newtxmath}
\usepackage{bm,times}
\usepackage{graphicx,epsfig}
\usepackage[colorlinks=true,breaklinks=true,linkcolor=blue,citecolor=blue,urlcolor=blue]{hyperref}

\def \tr{{\rm{Tr}}}

\def \d {\mathrm{d}}
\def \ket {\rangle}

\newcommand{\be}{\begin{eqnarray}}
\newcommand{\ee}{\end{eqnarray}}

\newcommand{\im}{\mathrm{i}}

\makeatletter
    
    \newcommand{\Rmnum}[1]{\expandafter\@slowromancap\romannumeral #1@}
\makeatother

\begin{document}
\author{Guangming Jiang}
\affiliation{College of Physical Science and Technology, Sichuan University, Chengdu 610064, China}

\author{Xiaohua Wu}
\email{wxhscu@scu.edu.cn}
\affiliation{College of Physical Science and Technology, Sichuan University, Chengdu 610064, China}

\author{Tao Zhou}
\email{taozhou@swjtu.edu.cn}
\affiliation{Quantum Optoelectronics Laboratory, School of Physical Science and Technology, Southwest Jiaotong University, Chengdu 610031, China}
\affiliation{Department of Applied Physics, School of Physical Science and Technology, Southwest Jiaotong University, Chengdu 611756, China}

\date{\today}

\title{Detecting the genuine multipartite two-way steerability with linear steering inequalities}

\begin{abstract}
According to the fundamental idea that a steering inequality can be constructed by just considering the measurements performed by Bob, and from the definitions of steering from Alice to Bob, a general scheme for designing two different kinds of linear steering inequalities (LSIs) is developed to detect the two-way steerability for bipartite system and the genuine multipartite two-way steerability for multipartite system, respectively. Besides the LSIs constructed from the known one-way criteria and the Bell operators, several other types of LSIs are also considered.
\end{abstract}

\pacs{03.65.Ud, 03.65.Ta}
\maketitle

\maketitle

\section{Introduction}
In 1930s, the concept of steering was introduced by Schr\"odinger~\cite{Sch} as a generalization of the Einstein-Podolsky-Rosen (EPR) paradox~\cite{Ein}. For a bipartite state, steering infers that an observer on one side can affect the state of the other spatially separated system by local measurements. In 2007, a standard formalism of quantum steering was developed by  Wiseman, Jones, and Doherty~\cite{Wiseman1}.  In quantum information processing, EPR steering can be defined as the task for a referee to determine whether one party shares entanglement with a second untrusted party~\cite{Wiseman1,JWD,sau}.  Quantum steering is a type of quantum nonlocality that is logically distinct from inseparability~\cite{Guhne,Horos} and Bell nonlocality~\cite{Brunner}.

A fundamental property is that steering is inherently asymmetric with respect to the observers~\cite{bowles,Midgley}, which is quite different from the quantum nonlocality and entanglement. The property of one-way steering has been predicted in a number of systems ~\cite{Bow, 37, 42} and demonstrated in several experimental configurations
~\cite{45,46, 48, 67}. Besides its foundational significance in quantum information theory, steering has been found useful in many applications. For examples, steering has a vast range of information-theoretic applications in one-sided device-independent scenarios where the party being steered has trust on his or her own quantum device while the other's device is untrusted, such as one-sided device-independent quantum key distribution~\cite{Bran}, advantage in subchannel discrimination~\cite{piani}, secure quantum teleportation~\cite{Reid1,He}, quantum communication~\cite{Reid1}, detecting bound entanglement~\cite{Mor}, one-sided device-independent randomness generation~\cite{law}, and one-sided device-independent self-testing of pure maximally as well as nonmaximally entangled state~\cite{supic}.

The detection and characterization of steering, have been widely discussed. In 1989, the variance inequalities violated with EPR correlations for continuous variable system were derived by Reid~\cite{eid}, and this was generalized to discrete variable systems~\cite{Caval}.  For a bipartite system,  EPR-steering inequalities were defined~\cite{can22}, where the violation of any such inequality implies steering. Following these works, further schemes have been proposed to signalize steering, for instance, the linear and nonlinear steering criteria~\cite{sau,wit,Pusey,Evan,mar,rut}, steering inequalities based on multiplicative variances~\cite{ReidRMD}, steering criteria from uncertainty relations~\cite{wa,schnee,Costaa,costab,jia,kri}, steering with Clauser-Horne-Shimony-Holt (CHSH)-like inequalities~\cite{Can3,Girdhar,cos,quan}, moment matrix approach~\cite{Kig,mo,chen00}, linear steering  inequality from the semidefinite program (SDP)  \cite{Can1}, steering criteria based on local uncertainty relations~\cite{Ji,Zhen}, and the universal steering criteria~\cite{Zhu}.

Besides the works focusing on the investigations of different steering criteria, some other works are devoted to determining the conditions under which it is possible to reveal steering and exploring how useful it is in practical applications~\cite{yuxiang}. Most works on demonstration of steering deals with optimal systems~\cite {sau,wit,Bennet,smith,weston,65,70}, and recently, the steering in multipartite system  has attracted much attention  and several approaches have been developed on this topic~\cite{he-reid,li-chen,csan,rmm,gmdrm, 51}. In this work, we shall introduce the definition of genuine  multipartite two-way steerability, and it can be viewed as a natural generalization of the two-way steerability which has a clear definition for the bipartite system~\cite{Can1,rmd}. A general protocol to design the sufficient criteria for detecting the genuine multipartite steerability with   linearly steering inequalities  (LSIs) will be developed. A special class of LSIs, which are constructed from the Bell operators, will be introduced, and furthermore, several other types of LSIs will also be considered.

The  content of this work is organized as follows. In Sec.~\ref{Sec2}, we give a brief review on the definitions of steering  and the most incompatible measurement. In Sec.~\ref{Sec3}, a detailed introduction to LSI for bipartite system is given there. In Sec.~\ref{Sec4}, we address the problem of detecting genuine multipartite two-way steerability with LSIs.  Some applications of the developed scheme are discussed in Sec.~\ref{Sec5}.  Finally, we end our work with a short conclusion.

\section{Preliminary}
\label{Sec2}

\subsection{Steering from Alice to Bob}
Before one can show how to demonstrate a state is steerable from Alice to Bob, some necessary conventions are required. First, Alice can perform $N$  measurements on her side, labelled by $\mu=1,2,...,N$, each having $m$ outcomes $a=0,1,...,m-1$, and the measurements are denoted by $\hat{\Pi}^{a}_{\mu}$, $\sum_{a=0}^{m-1}\hat{\Pi}^{a}_{\mu}=I_d $, with $I_d$ the identity operator for the local $d$-dimensional Hilbert space. For a bipartite state $W$,  the unnormalized post-measurement states prepared for Bob are given by
\begin{equation}
\label{df}
\tilde{\rho}_{\mu}^a=\mathrm{Tr}_{A}[(\hat{\Pi}^{a}_{\mu}\otimes I_d) W].
\end{equation}
The set of unnormalized states, $\{\tilde{\rho}^{a}_{\mu}\}$, is usually called an \emph{assemblage}.

In 2007, Wiseman, Jones, and Doherty formally defined quantum steering as the possibility of remotely generating ensembles that could not be produced by a~\emph{local hidden states} (LHS) model~\cite{Wiseman1}. An LHS model refers to the case where a source sends a classical message $\xi$ to one of the two parties, say, Alice, and a corresponding quantum state $\rho_{\xi}$ to the other party, say Bob.  Given that  Alice decides to performs the $\mu$th measurement, the variable $\xi$ instructs the output $a$ of Alice's apparatus with the probability $\mathfrak{p}(a\vert\mu,\xi)$. The variable $\xi$ is usually chosen according to a probability distribution $\Omega(\xi)$ and can also be interpreted as a local hidden variable (LHV). Bob does not have access to the classical variable $\xi$, and his final assemblage is composed by the LHS model
\begin{equation}
\label{tilderho}
\tilde{\rho}^{a}_{\mu}=\int d\xi\Omega(\xi)\mathfrak{p}(a\vert\mu,\xi)\rho_{\xi}
\end{equation}
with $\int d\xi\Omega(\xi)=1$.

In this paper, the definition of steering is directly cited from the review article~\cite{Can1}: An assemblage is said to demonstrate steering if it does not admit a decomposition of the form in Eq.~\eqref{tilderho}.  Furthermore, a quantum state $W$ is said to be steerable from Alice to Bob if the experiments in Alice's part produce an assemblage that demonstrate steering. On the contrary, an assemblage is said to be LHS if it can be written as in Eq.~\eqref{tilderho}, and a quantum state is said to be unsteerable  if an LHS assemblage is always generated for all local measurements.

Via a similar argument, one can give  a definition of steering from Bob to Alice. A state is said to be two-way steerable if it is steerable both from Alice to Bob and from Bob to Alice.

\subsection{Most incompatible measurements}
A set of measurements $\{\hat{M}^a_{\mu}\}$ is compatible, if there exists a set of positive-operator-valued-measures (POVMs) $\{\hat{M}_{\lambda}\}$ such that $\hat{M}^a_{\mu}=\sum_{\lambda}\pi(\lambda) p(a\vert\mu,\lambda)\hat{M}_{\lambda}$ for all $a$ and $\mu$, where $\pi(\lambda)$ and $p(a\vert\mu,\lambda)$ are the probability distributions. If such measurements are performed by Alice, from  Eq.~\eqref{df}, the assemblage $\{ \tilde{\rho}^{a}_{\mu}\}$ will admit an LHS model, $\tilde{\rho}^{a}_{\mu}=\sum_{\lambda}\pi(\lambda) p(a\vert\mu,\lambda){\rho}_{\lambda}$, with the LHS states ${\rho}_{\lambda}=\mathrm{Tr}_{A}[(\hat{M}_{\lambda}\otimes I_d) W].$

By introducing the critical visibility, a quantity used to characterize the white-noise robustness of an assemblage, Bavareso~\emph{et. al.} recently addressed the problem of finding the most incompatible measurements when $N$ and $m$ are fixed~\cite{Bava}. Consider a depolarizing map $\varepsilon_{\eta}$ acting on the Hermitian operator $\hat{A}$ of a $d$-dimensional Hilbert space $\mathcal{H}_d$, $\varepsilon_{\eta}(\hat{A})=\eta\hat{A}+(1-\eta)\mathrm{Tr}(\hat{A})I_d/d$, and the critical visibility is defined as $\eta(\hat{M}^a_{\mu})\equiv\max[\eta\vert \{\varepsilon_{\eta}(\hat{M}^a_{\mu})\}_{a,\mu}\in\mathcal{LHS} ]$, where $\mathcal{LHS}$ is a set of assemblages that admit an LHS model. The quantity $\eta(\hat{M}^a_{\mu})$ is the exact value of $\eta$ above which the assemblage $\{\varepsilon_{\eta}(\hat{M}^a_{\mu}\}_{a,\mu}\}$ no longer admits an LHS model, and when $N$ and $m$ are fixed, the optimal critical visibility, $\eta^*({N, m})$ is defined as $\eta^*(N, m)\equiv\min_{\{\hat{M}^a_{\mu}\}_{a,\mu}}\eta(\hat{M}^a_{\mu}).$

It is known that a set of mutually unbiased bases (MUBs) consists of two or more orthonormal basis $\{\vert\phi_{x}^a\rangle\}$ in a $d$-dimensional Hilbert space satisfying
\begin{equation}
\label{MUB1}
\left\vert\langle\phi^a_x\right\vert\phi^b_y\rangle\vert^2=\frac{1}{d},~\forall a,b\in\{0,1,...,d-1\},~x\neq y,
\end{equation}
for all $x$ and $y$~\cite{mubs}, and two results can be obtained from the MUBs for two-dimensional systems~\cite{Bava}
\begin{equation}
\label{MUBs}
\eta^*(N=2,m=2)=\frac{1}{\sqrt{2}},\ \ \eta^*(N=3,m=2)=\frac{1}{\sqrt{3}},
\end{equation}
which are useful in the following. For the Pauli matrices $\sigma_j$ $(j=x,y,z)$, their eigenvectors $\vert\psi^{a}_j\rangle$, where $\vert\psi^{a}_j\rangle=(I_2+(-1)^a\sigma_j)/2$ with
$a\in\{0,1\}$, form a set of the MUBs. According to Eq.~\eqref{MUBs}, the set of two measurements, $\{\varepsilon_{\eta}(\vert\psi^{a}_j\rangle\langle\psi^a_j\vert)\}$ with ${j=x,y}$ (or $j=x,z$), is compatible if $\eta\leqslant1/\sqrt{2}$. The  set of three measurements, $\{\varepsilon_{\eta}(\vert\psi^{a}_j\rangle\langle\psi^a_j\vert)\}$ with ${j=x,y,z}$, is also compatible if $\eta\leqslant1/\sqrt{3}$.

\section{Linear steering inequalities for bipartite system}
\label{Sec3}
The LSIs originate from the works in Refs.~\cite{can22,sau,Joness}. To discuss the one-way steering from Alice to Bob, one may construct a criterion which only depends on the measurements performed by Bob. Besides the property that the LSIs  can work even when the state is unknown, they also have a deep relation  with the compatible  measurement: If a one-way LSI is violated, the state is steerable from Alice to Bob and the measurements performed by Alice are also verified to be incompatible~\cite{quint,ula,UULA,Kiukas,Wu,WU2}. In this section,  we shall develop a general scheme to construct the LSIs for detecting the two-way steerability of the bipartite system.

\subsection{ Sufficient criteria for steering}
For a bipartite system $\mathcal{H}_{\mathrm{A}}\otimes \mathcal{H}_\mathrm{B}$, the POVMs $\hat{\Pi}^a_\mu$ ($\hat{M}^b_\nu$) can be introduced for the local Hilbert space $\mathcal{H}_\mathrm{A}$ ($\mathcal{H}_\mathrm{B}$). Certainly, $\sum_{a} \hat{\Pi}^a_\mu=I_{\mathrm{A}}$ and $\sum_{b} \hat{M}^b_\nu=I_{\mathrm{B}}$, where $I_{\mathrm{A}}$ and $I_{\mathrm{B}}$ are the identity operator on $\mathcal{H}_\mathrm{A}$ and $\mathcal{H}_\mathrm{B}$, respectively. In general, one can introduce a Hermitian operator
\begin{equation}
\label{defining-H}
\hat{H}=\sum_{\mu}\sum_{\nu}\sum_a\sum_b c(ab\vert\mu\nu)\hat{\Pi}^a_{\mu}\otimes \hat{M}^b_{\nu}
\end{equation}
 where the coefficients $c(ab\vert \mu\nu)$ are real values.

To discuss the steering from Alice to Bob ($\mathrm{A}\rightarrow \mathrm{B}$), the operator $\hat{H}$ can be rewritten as
$\hat{H}=\sum_{\mu}q_{\mu}\sum_{a}\hat{\Pi}^{a}_{\mu}\otimes \hat{F}^{a}_{\mu}$.
It can be understood as that: Assume the probability of the $\mu$th measurement performed by Alice is $q_{\mu}$, $\sum_{\mu=1}^N q_{\mu}=1$, and the conditional state $\tilde{\rho}^{a}_{\mu}$ on Bob's side are  measured with a set of Hermitian operators $\{\hat{F}^a_{\mu}\}$, $ \hat{F}^a_{\mu}=(\hat{F}^a_{\mu})^{\dagger}$. For the $\mu$th run of experiment,  a quantity $\beta_{\mu}$ can be defined, $\beta_{\mu}=\sum_{a=0}^{m-1}\mathrm{Tr}(\tilde{\rho}^{a}_{\mu}\hat{F}^a_{\mu})$. Let $\langle A\otimes B  \rangle\equiv\mathrm{Tr}\left[(A\otimes B)W\right]$ be the expectation value of the operator $A\otimes B$, and in experiment,  $\beta_{\mu}$ can be measured as
\begin{equation}
\beta_{\mu}=\sum_{a=0}^{m-1}\left\langle \hat{\Pi}^a_{\mu}\otimes \hat{F}^a_{\mu}\right\rangle.
\end{equation}
The averaged expectation of the set of operators $\{\hat{\Pi}^a_{\mu}\otimes \hat{F}^a_{\mu}\}$ can be defined,
$\beta\equiv\sum_{\mu=1}^N q_{\mu}\beta_{\mu}$. If the assemblage $\{\tilde{\rho}^a_{\mu}\}$ introduced in Eq.~\eqref{df} has an LHS decomposition in Eq.~\eqref{tilderho}, one has $\mathrm{Tr}(\tilde{\rho}^{a}_{\mu}\hat{F}^a_{\mu})=\int d\xi\Omega(\xi)\mathfrak{p}(a\vert\mu,\xi)\mathrm{Tr}(\rho_{\xi}\hat{F}^{a}_{\mu})$, and an averaged expectation can be  introduced
\begin{equation}
\beta^{\mathrm{LHS}}_{\mathrm{avg}}\equiv\sum_{\mu=1}^N\sum_{a=0}^{m-1}q_{\mu}
\int d\xi\Omega(\xi)\mathfrak{p}(a\vert\mu,\xi)\mathrm{Tr}(\rho_{\xi}\hat{F}^{a}_{\mu}).
\end{equation}
Formally, $\beta^{\mathrm{LHS}}_{\mathrm{avg}}=\int d\xi\Omega(\xi)\mathrm{Tr}[\rho_{\xi}\hat{H}(\xi)]$, with
\begin{equation}
\label{rhobar}
\hat{H}(\xi)=\sum_{\mu=1}^N\sum_{a=0}^{m-1}q_{\mu}\mathfrak{p}(a\vert\mu,\xi)\hat{F}^{a}_{\mu},
\end{equation}
which can be introduced in an operational way: First, write down an operator $\hat{H}=\sum_{\mu}\sum_{a}q_{\mu}\hat{\Pi}^a_{\mu}\otimes\hat{F}^a_{\mu}$, and then $\hat{H}(\xi)$ will be obtained by replacing each operator $\hat{\Pi}^a_{\mu}$ with the probability $\mathfrak{p}(a\vert\mu,\xi)$, which is interpreted as the predetermined value of $\hat{\Pi}^a_{\mu}$ in the LHV model. Obviously, $\hat {H}(\xi)$ is a Hermitian operator, and can be expanded as $\hat{H}(\xi)=\sum_{\nu}\lambda_{\nu}\vert \lambda_\nu\rangle\langle \lambda_{\nu}\vert$, with $\lambda_{\nu}$ the eigenvalues and $\vert \lambda_{\nu}\rangle$ the corresponding eigenvectors. Defining
\begin{equation}
\vert\hat{H}(\xi)\vert^{\mathrm{max}}\equiv\lambda^{\max}=\max_{\mathfrak{p}(a\vert\mu,\xi)}\max_{\mu}\{\lambda_{\mu}\},
\end{equation}
and together with the facts $\mathrm{Tr}[\rho_{\xi}\hat{H}(\xi)]\leqslant\lambda^{\max}$ and $\int \Omega(\xi)d\xi=1$, one can conclude that $ \vert\hat{H}\vert^{\mathrm{max}}$ is an upper bound of $\beta^{\mathrm{LHS}}_{\mathrm{avg}}$, say, $\vert\hat{H}\vert^{\mathrm{max}}\geqslant \beta^{\mathrm{LHS}}_{\mathrm{avg}}$.
From the definition of unsteerable states, the assemblage resulted from the unsteerable state always admits an LHS model. Therefore, $\vert\hat{H}\vert^{\mathrm{max}}$ can also be interpreted as the upper bound of the averaged expectation, which can be obtained from the unsteerable  states, if the measurement on Bob's side has been fixed as $\{ q_{\mu},\hat{F}^a_{\mu}\}$. To emphasize this property of $\vert\hat{H}\vert^{\mathrm{max}}$, we call it as the steering threshold (ST)
and denote it by the symbol $\beta_{\mathrm{ST}}^{\mathrm{A}\rightarrow \mathrm{B}}$ hereafter,
\begin{equation}
\label{NST+}
\beta_{\mathrm{ST}}^{\mathrm{A}\rightarrow \mathrm{B}}(\{q_{\mu},\hat{F}^a_{\mu}\})=
\max_{\vert\phi\rangle}\max_{\mathfrak{p}(a\vert\mu,\xi)}
\langle\phi\vert\hat{H}(\xi)\vert\phi\rangle.
\end{equation}

From $\hat{H}(\xi)=\sum_{\nu}\lambda_{\nu}\vert \lambda_\nu\rangle\langle \lambda_{\nu}\vert$, one can also define
\begin{equation}
 \vert\hat{H}(\xi)\vert^{\mathrm{min}}\equiv\lambda^{\min}
 =\min_{\mathfrak{p}(a\vert\mu,\xi)}\min_{\mu}\{\lambda_{\mu}\}.
 \end{equation}
 With the facts that $\mathrm{Tr}[\rho_{\xi}\hat{H}(\xi)]\geqslant\lambda^{\min}$
 and $\int \Omega(\xi)d\xi=1$, one can conclude that
$ \vert\hat{H}(\xi)\vert^{\mathrm{min}}$ is a lower
bound of $\beta^{\mathrm{LHS}}_{\mathrm{avg}}$, say,
$ \vert\hat{H}(\xi)\vert^{\mathrm{min}}\leqslant \beta^{\mathrm{LHS}}_{\mathrm{avg}}$.
Another type of steering threshold, which is denoted
 by the symbol $\gamma_{\mathrm{ST}}^{\mathrm{A}\rightarrow \mathrm{B}}$, can be introduced
\begin{equation}
\label{NST-}
\gamma_{\mathrm{ST}}^{\mathrm{A}\rightarrow \mathrm{B}}(\{q_{\mu},\hat{F}^a_{\mu}\})=
\min_{\vert\phi\rangle}\min_{\mathfrak{p}(a\vert\mu,\xi)}\langle\phi\vert\hat{H}(\xi)\vert\phi\rangle.
\end{equation}

Now, a one-way  LSI for $\mathrm{A}\rightarrow \mathrm{B}$ can be defined
\begin{equation}
\label{LSI}
\gamma^{\mathrm{A}\rightarrow \mathrm{B}}_{\mathrm{ST}}(\{q_{\mu},\hat{F}^a_{\mu}\})
\leqslant \langle\hat{H}\rangle\leqslant\beta_{\mathrm{ST}}^{\mathrm{A}\rightarrow \mathrm{B}}(\{q_{\mu},\hat{F}^a_{\mu}\}).
\end{equation}
Since each of the following two conditions: (a) The state is steerable from Alice to Bob, and (b) The set of measurements $\{\hat{\Pi}^a_{\mu}\}$ performed by Alice is incompatible, is necessary so that the assemblage $\{{\tilde{\rho}}^{a}_{\mu}\}$ does not admit
an LHS  model, one may conclude that the violation of the steering inequality,
is a sufficient condition for Bob to make the statements (a) and (b).

To show whether a state $W$ is steerable from Alice to Bob, the extremal value of the averaged expectation should be considered. First, let us consider the probabilistic  model, where for the $\mu$th measurement $\{\hat{\Pi}^{a}_{\mu}\}$,
\begin{equation}
0\leqslant\mathfrak{p}(a\vert\mu,\xi)\leqslant 1,~\sum_{a=0}^{m-1}\mathfrak{p}(a\vert\mu,\xi)=1.
\end{equation}
A quantity $f_{\mu}(\phi)=\langle\phi\vert\sum_{a=0}^{m-1} \mathfrak{p}(a\vert\mu,\xi)\hat{F}^{a}_{\mu}\vert\phi\rangle$ can be introduced.  For a fixed $\vert\phi\rangle$, one can select out an operator $\hat{F}^{\tilde{a}}_{\mu}$ from the set of operators $\{\hat{F}^{a}_{\mu}\}_{a=0}^{m-1}$ with the constraint $\langle\phi \vert \hat{F}^{\tilde{a}}_{\mu}\vert\phi\rangle \geqslant\langle\phi \vert \hat{F}^{{a}}_{\mu}\vert\phi\rangle$, $\forall a\in\{0,1,...,m-1\}$. The maximum value of $f_{\mu}(\phi)$, $f^{\max}_{\mu}(\phi)=\langle\phi\vert\hat{F}^{\tilde{a}}_{\mu}\vert\phi\rangle,$
can be obtained with the optimal choice of the probabilities $\{\mathfrak{p}(a\vert\mu,\xi)\}$
\begin{equation}
\label{optpro}
\mathfrak{p}^\star(a\vert\mu,\xi)=\delta_{a\tilde{a}}.
\end{equation}
The one-way steering threshold $\beta_{\mathrm{ST}}^{\mathrm{A}\rightarrow \mathrm{B}}$
 can be rewritten as
$\beta_{\mathrm{ST}}^{\mathrm{A}\rightarrow \mathrm{B}}=\max_{\vert\phi\rangle}\sum_{\mu=1}^N q_{\mu}f^{\max}_{\mu}(\phi)$.

Via a similar argument, if   an operator $\hat{F}^{\tilde{a}}_{\mu}$  is selected out from  the set  $\{\hat{F}^{a}_{\mu}\}_{a=0}^{m-1}$ with the constraint $\langle\phi \vert \hat{F}^{\tilde{a}}_{\mu}\vert\phi\rangle \leqslant\langle\phi \vert \hat{F}^{{a}}_{\mu}\vert\phi\rangle$, $\forall a\in\{0,1,...,m-1\}$, the minimum value of $f_{\mu}(\phi)$, $f^{\min}_{\mu}(\phi)=\langle\phi\vert\hat{F}^{\tilde{a}}_{\mu}\vert\phi\rangle$, can also be obtained with the optimal choice in Eq. \eqref{optpro}. The steering threshold $\gamma_{\mathrm{ST}}^{\mathrm{A}\rightarrow \mathrm{B}}$ can be rewritten as
$\gamma_{\mathrm{ST}}^{\mathrm{A}\rightarrow \mathrm{B}}=\min_{\vert\phi
\rangle}\sum_{\mu=1}^N q_{\mu}f^{\min}_{\mu}(\phi)$.

From the optimal choice of $\{\mathfrak{p}(a\vert\mu,\xi)\}$, it is shown that steering thresholds remain unchanged if a deterministic model is applied, 
\begin{equation}
\mathfrak{p}(a\vert\mu,\xi)\in\{0,1\},~~\sum_{a=0}^{m-1}\mathfrak{p}(a\vert\mu,\xi)=1.
\end{equation}
So, another convenient way to derive the one-way LSI can be constructed, shown in the following. Considering the steering from Alice to Bob where $\hat{H}(\xi)=\sum_{\mu}\sum_{a} q_{\mu}\mathfrak{p}(a\vert\mu,\xi)\hat{F}^a_{\mu}$, and within the deterministic model above, one may introduce  a series of Hermitian operators
\begin{equation}
\label{hhhh}
\hat{H}_{k_1,k_2,...,k_N}=\sum_{\mu=1}^N q_{\mu}\hat{F}_{\mu}^{k_{\mu}},
\end{equation}
where $k_\mu\in\{0,1,...,m-1\}$ for all $\mu=1,2,...,N$, and there are totally $m^N$ operators of such kind. With the denotations $\vert  \hat{H}_{k_1,k_2,...,k_N}\vert^{\mathrm{\max}}=\max_{\vert\phi\rangle}\langle\phi\vert \hat{H}_{k_1,k_2,...,k_N}\vert\phi\rangle$, and $\vert  \hat{H}_{k_1,k_2,...,k_N}\vert^{\mathrm{\min}}=\min_{\vert\phi\rangle}\langle\phi\vert \hat{H}_{k_1,k_2,...,k_N}\vert\phi\rangle$, the steering thresholds can be expressed as
\begin{eqnarray}
\label{beta1}
\beta_{\mathrm{ST}}^{\mathrm{A}\rightarrow \mathrm{B}}&=&\max_{\{k_j\}}\left\{\vert  \hat{H}_{k_1,k_2,...,k_N}\vert^{\mathrm{\max}}\right\},\\
\gamma_{\mathrm{ST}}^{\mathrm{A}\rightarrow \mathrm{B}}&=&\min_{\{k_j\}}\left\{\vert  \hat{H}_{k_1,k_2,...,k_N}\vert^{\mathrm{\min}}\right\}.
\end{eqnarray}

As an illustration, let us consider a two-result  case as a specific example. For the $\mu$th run of experiment, Alice performs the measurement $\hat{\Pi}^0_{\mu}$ and $\hat{\Pi}^1_{\mu}$, and $\sum _{a=0}^{1}\hat{\Pi}_{\mu}^a=I_d$. Meanwhile, the measurements on Bob's side are fixed as $\hat{F}^0_{\mu}=\hat{F}_{\mu},\hat{F}^1_{\mu}=-\hat{F}_{\mu}$. Furthermore, we assume that the experiment is realized in an equal-weighted way, $q_{\mu}=1/N$. From Eq.~\eqref{hhhh}, one can have $\hat{H}_{k_1,k_2,...,k_N}=\sum_{j=1}^{N} (-1)^{k_j}\hat{F}_{\mu}/N, \forall k_j\in\{0,1\}$, and with the experiment data $\beta=\sum_{\mu}q_{\mu}\langle (\hat{\Pi}^{0}_{\mu}-\hat{\Pi}^{1}_{\mu})\otimes \hat{F}_{\mu}\rangle$, the LSI takes the form:
\begin{equation}
\sum_{\mu}q_{\mu}\left\langle (\hat{\Pi}^{0}_{\mu}-\hat{\Pi}^{1}_{\mu})\otimes
\hat{F}_{\mu}\right\rangle\leqslant q_1\max_{\{k_j\}}\left\{\vert\sum_{j=1}^{N} (-1)^{k_j}\hat{F}_{\mu}\vert^{\max}\right\}.\nonumber
\end{equation}
The result in Ref.~\cite{sau} can be recovered here. One may note that in the above inequality, the probability $q_{\mu}$ can  be absorbed in the operator $\hat{F}_{\mu}$. In the following, we usually  work with $q_\mu\hat{F}_{\mu}\rightarrow \hat{F}_{\mu}$, $\forall\mu\in\{1,2,..., N\}$.

To discuss the steering from Bob to Alice ($ \mathrm{A}\leftarrow \mathrm{B })$, the operator $\hat{H}$ can be rewritten as $\hat{H}=\sum_{\nu}\sum_{b}\hat{F}^b_{\nu}\otimes \hat{M}^b_{\nu}$.
Then,  by replacing each $\hat{M}^b_{\nu}$ with $\mathfrak{p}(b\vert\nu,\xi)$, an operator $\hat{H}(\xi)$ is introduced as $\hat{H}(\xi)=\sum_{\nu}\sum_{b}\mathfrak{p}(b\vert\nu,\xi)\hat{F}^b_{\nu}$. Formally, another one-way LSI can be obtained, $\gamma^{\mathrm{A}\leftarrow \mathrm{B}}_{\mathrm{ST}}\leqslant \langle\hat{H}\rangle\leqslant\beta_{\mathrm{ST}}^{\mathrm{A}\leftarrow \mathrm{B}}$, the violation of which shows that the state is steerable from Bob to Alice.

With the two one-way LSIs introduced above, the steering threshold $\beta_{\mathrm{ST}}$ and $\gamma_{\mathrm{ST}}$ can be defined as
\begin{eqnarray}
\label{beta}
\beta_{\mathrm{ST}}&=&\max\{\beta_{\mathrm{ST}}^{\mathrm{A}\rightarrow \mathrm{B}},\beta_{\mathrm{ST}}^{\mathrm{A}\leftarrow \mathrm{B}}\},\\
\label{gamma}
\gamma_{\mathrm{ST}}&=&\min\{\gamma_{\mathrm{ST}}^{\mathrm{A}\rightarrow \mathrm{B}},\gamma_{\mathrm{ST}}^{\mathrm{A}\leftarrow \mathrm{B}}\}.
\end{eqnarray}
The violation of the  LSI, $\gamma_{\mathrm{ST}}\leqslant\langle\hat{H}\rangle\leqslant\beta_{\mathrm{ST}}$, indicates that the state is two-way steerable. Obviously,   to  construct a two-way steering LSI,  a pair of one-way LSIs, $\gamma^{{\mathrm{A}}\rightarrow\mathrm{B}}_{\mathrm{ST}}\leqslant\langle\hat{H}\rangle\leqslant\beta^{\mathrm{A}\rightarrow\mathrm{B}}_{\mathrm{ST}}$ and $\gamma^{{\mathrm{A}}\leftarrow\mathrm{B}}_{\mathrm{ST}}
\leqslant\langle\hat{H}\rangle\leqslant\beta^{\mathrm{A}\leftarrow\mathrm{B}}_{\mathrm{ST}}$,
are needed.

Here, it should be mentioned that the above inequality, $\gamma_{\mathrm{ST}}\leqslant \langle\hat{H}\rangle\leqslant\beta_{\mathrm{ST}}$, can be divided into two independent ones: (1) $\gamma_{\mathrm{ST}}\leqslant \langle\hat{H}\rangle$ and (2) $\langle\hat{H}\rangle\leqslant\beta_{\mathrm{ST}}$, and a state is verified to be two-way steerable, if one of the two inequalities is violated. In the following, two examples are given where both $\gamma_{\mathrm{ST}}$ and $\beta_{\mathrm{ST}}$ are  presented. For the rest of examples, for simplicity,  only the type (2) inequality shall be considered.

In the present work, for a local $d$-dimensional Hilbert space, we usually refer to the special POVM, $\hat{\Pi}^0+\hat{\Pi}^1=\mathrm{I}_d$, as the two-result measurement, while an operator $\hat{A}$ is a $2$-value operator if it has just two eigenvalues $\pm 1$: $\hat{A}\vert m,+\rangle=\vert m,+\rangle$, and  $\hat{A}\vert n,-\rangle=-\vert n,-\rangle$. For such an operator, the corresponding two-result measurement can be defined as $\hat{\Pi}^0=\sum_m \vert m,+\rangle\langle m,+\vert$, $\hat{\Pi}^1=\sum_n \vert n,-\rangle\langle n,-\vert$, from which, one can obtain $\hat{\Pi}^0-\hat{\Pi}^1=\hat{A}$, and $\hat{\Pi}^0+\hat{\Pi}^1=I_d$, with $I_d$ the identity operator for the local $d$-dimensional Hilbert space where $\hat{A}$ is defined. To discuss the steering from Alice to Bob, one can introduce the denotation
\begin{equation}
\label{ddddd}
\mathfrak{d}(\mu\vert\xi)=\mathfrak{p}(0\vert\mu,\xi)-\mathfrak{p}(1\vert\mu,\xi),
\ \ \mathfrak{d}(\mu\vert\xi)\in\{\pm1\}.
\end{equation}
Under the condition that the operator $\hat{H}$ is  expanded as $\hat{H}=\sum_{\mu}\hat{A}_{\mu}\otimes \hat{F}_{\mu}$ with $\hat{A}_{\mu}=\hat{\Pi}^0_{\mu}-\hat{\Pi}^{1}_{\mu}$, $\hat{H}(\xi)$ can be obtained from $\hat{H}$ by simply replacing each $\hat{A}_{\mu}$ with $\mathfrak{d}(\mu\vert\xi)$, say, $\hat{H}(\xi)=\sum_{\mu}\mathfrak{d}(\mu\vert\xi)\hat{F}_{\mu}$.

\subsection{Constructing two-way criteria  from the known one-way LSIs}
In previous works, a series of one-way LSIs from Alice to Bob have been constructed, and in this section, the construction of two-way LSIs from these known criteria will be discussed.

Let us start from the two well-known inequalities:
\begin{eqnarray}
\label{lsi1}\langle\hat{A}_1\otimes \sigma_x\rangle+\langle\hat{A}_2\otimes\sigma_z\rangle\leqslant&&\sqrt{2},\\
\label{lsi2}
\langle\hat{A}_1\otimes \sigma_x\rangle+\langle\hat {A}_2\otimes\sigma_z\rangle+\langle\hat{A}_3\otimes\sigma_y\rangle\leqslant&&\sqrt{3}.
\end{eqnarray}
where $\hat{A}_{\mu}$ are arbitrary 2-value operators. The two inequalities above firstly appeared in Ref.~\cite {can22}, and recently, it was demonstrated that the criteria can be returned from the semidefinite program~\cite{Can1}. Consider the LSI in Eq.~\eqref{lsi1} at first. For the steering from Bob to Alice, with Eq.~\eqref{ddddd}, there are four possible $\hat{H}(\xi)$: $\hat{H}(\xi)_{\pm\pm}=\pm\hat{A}_1\pm\hat{A}_2$. Take the $\hat{H}_{++}$ as an example, its maximum eigenvalue will be calculated with the
following rules: (1) For arbitrary operators $\hat{A}_{\mu}$ ($\mu=1,2$), there is $\vert \hat{A}_1+\hat{A}_2\vert^{\max}\leqslant\vert \hat{A}_1\vert^{\max}+\vert \hat{A}_2\vert^{\max}$; and (2) If $\hat{A}_{\mu}$ are known operators, one can perform a standard calculation, say,  $\vert\hat{A}_1+\hat{A}_2\vert^{\max}=\max_{\vert\phi\rangle}\langle\phi\vert\hat{A}_1+\hat{A}_2\vert\phi\rangle$. These calculation  rules can be easily generalized for the case with more than two operators.  Therefore, for arbitrary operators $\hat{A}_{\mu}$ ($\mu=1,2$), one can have $\vert\hat{H}(\xi)_{\pm\pm}\vert^{\max}\leqslant 2$ and $\beta^{\mathrm{A}\leftarrow \mathrm{B}}_{\mathrm{ST}}=2$. Obviously, the one-way LSI from Bob to Alice, $\langle \hat{H}\rangle \leqslant \beta^{\mathrm{A}\leftarrow \mathrm{B}}_{\mathrm{ST}}=2,$ cannot be violated.   To derive a lower value of $ \beta^{\mathrm{A}\leftarrow \mathrm{B}}_{\mathrm{ST}}$, some additional constraints for the operators $\hat{A}_{\mu}$ are required. For example, a type of constraints can be introduced in the following:
\begin{equation}
\label{contraint1}
\hat{A}_1=U\sigma_xU^{\dagger}, \hat{A}_2=U\sigma_zU^{\dagger},
\end{equation}
with $U$ an arbitrary two-dimensional unitary operator. Based on it, there is $\beta^{\mathrm{A}\leftarrow \mathrm{B}}_{\mathrm{ST}}=\sqrt{2}$. Now, one can have a conclusion: For the operator $\hat{H}=\hat{A}_1\otimes \sigma_x+\hat{A}_2\otimes \sigma_z$ used in Eq.~\eqref{lsi1}, the steering threshold $\beta^{\mathrm{A}\leftarrow \mathrm{B}}_{\mathrm{ST}}=2$, which  holds for the unknown 2-value operators, will take a lower value if some additional constraints have been introduced.

Return to the operator in Eq.~\eqref{defining-H}, it is obvious that the LSIs can be constructed by taking the coefficients $c(ab\vert \mu\nu)$ as free parameters. As an example, one may introduce an operator
\begin{equation}
\label{lsi11}
\hat{H}(\theta)=\sin\theta \hat{A}_1\otimes \sigma_x+\cos\theta \hat{A}_2\otimes \sigma_z,
\end{equation}
as a simple generalization of the original one $\hat{H}=\hat{A}_1\otimes \sigma_x+\hat{A}_2\otimes \sigma_z$. For the steering from Alice to Bob, there are four possible $\hat{H}({\theta,\xi})$, $\hat{H}({\theta,\xi})_{\pm\pm}=\pm\sin\theta \sigma_x\pm\cos\theta\sigma_z$. It is easy to see that $\vert\pm\sin\theta \sigma_x\pm\cos\theta\sigma_z\vert^{\max}=1$, and $\beta^{\mathrm{A}\rightarrow \mathrm{B}}_{\mathrm{ST}}=1$. With the constraint in Eq.~\eqref{contraint1}, one can obtain $\beta^{\mathrm{A}\leftarrow \mathrm{B}}_{\mathrm{ST}}=1$. Thus, the two-way steering LSI is known as
\begin{equation}
\label{glsi1}
\sin\theta \langle\hat{A}_1\otimes \sigma_x\rangle+\cos\theta \langle\hat{A}_2\otimes \sigma_z\rangle\leqslant 1,
\end{equation}
By letting $\cos\theta=\sin\theta=\sqrt{2}/2$, the LSI in Eq. \eqref{lsi1} is recovered.

In experiment, if the correlations, $\langle\hat{A}_1\otimes \sigma_x\rangle$ and  $\langle\hat{A}_2\otimes \sigma_z\rangle$, have been decided,  the optimal choice for $\theta$ satisfies
\begin{eqnarray}
\label{theta}
\sin\theta&=&\frac{\langle\hat{A}_1\otimes \sigma_x\rangle}{\sqrt{\langle\hat{A}_1\otimes \sigma_x\rangle^2+
\langle\hat{A}_2\otimes \sigma_z\rangle^2}},\nonumber\\
\cos\theta&=&\frac{\langle\hat{A}_2\otimes \sigma_z\rangle}{\sqrt{\langle\hat{A}_1\otimes \sigma_x\rangle^2+
\langle\hat{A}_2\otimes \sigma_z\rangle^2}}.\nonumber
\end{eqnarray}
Finally, a two-way steering criterion is arrived at,
\begin{equation}
\sqrt{\langle\hat{A}_1\otimes \sigma_x\rangle^2+
\langle\hat{A}_2\otimes \sigma_z\rangle^2}\leqslant 1.\nonumber
\end{equation}

Similarly, one can design a two-parameters operator,
\begin{eqnarray}
\hat{H}(\theta,\phi)&=&
\sin\theta\cos\phi\hat{A}_1\otimes \sigma_x\nonumber\\ &+&\cos\theta \hat{A}_2\otimes \sigma_z+
\sin\theta\sin\phi\hat{A}_3\otimes \sigma_y,\nonumber
\end{eqnarray}
as the generalization of the operator in Eq.~\eqref{lsi2}. 
For the steering from Alice to Bob, there are eight possible operators
$\hat{H}_{\pm\pm\pm}$, $\hat{H}_{\pm\pm\pm}=\pm\sin\theta\cos\phi \sigma_x \pm\cos\theta \sigma_z\pm \sin\theta\sin\phi \sigma_y$, within the deterministic model. It is easy to get $\vert \hat{H}_{\pm\pm\pm}\vert^{\max}=1$ and $\beta^{\mathrm{A}\rightarrow \mathrm{B}}_{\mathrm{ST}}=1$.
By introducing the additional constraints
\begin{equation}
\label{contraint2}
\hat{A}_1=U\sigma_xU^{\dagger}, \hat{A}_2=U\sigma_zU^{\dagger},\hat{A}_3=U\sigma_yU^{\dagger},
\end{equation}
 there should be $\beta^{\mathrm{A}\leftarrow \mathrm{B}}_{\mathrm{ST}}=1$.  The  two-way steering LSI is arrived at,
$\langle\hat{H}(\theta,\phi)\rangle \leqslant 1$.
 With the optimal choices,
\begin{eqnarray}
\sin\theta&=&\frac{\sqrt{\langle\hat{A}_1\otimes \sigma_x\rangle^2+\langle\hat{A}_3\otimes \sigma_y\rangle^2}}{\sqrt{\langle\hat{A}_1\otimes \sigma_x\rangle^2+
\langle\hat{A}_2\otimes \sigma_z\rangle^2+\langle\hat{A}_3\otimes \sigma_y\rangle^2}},\nonumber\\
\cos\theta&=&\frac{\langle\hat{A}_2\otimes \sigma_z\rangle}{\sqrt{\langle\hat{A}_1\otimes \sigma_x\rangle^2+
\langle\hat{A}_2\otimes \sigma_z\rangle^2+\langle\hat{A}_3\otimes \sigma_y\rangle^2}},\nonumber\\
\cos\phi&=&\frac{\langle\hat{A}_1\otimes \sigma_x\rangle}{\sqrt{\langle\hat{A}_1\otimes \sigma_x\rangle^2+\langle\hat{A}_3\otimes \sigma_y\rangle^2}},\nonumber\\
\sin\phi&=&\frac{\langle\hat{A}_3\otimes \sigma_y\rangle}{\sqrt{\langle\hat{A}_1\otimes \sigma_x\rangle^2+\langle\hat{A}_3\otimes \sigma_y\rangle^2}},\nonumber
\end{eqnarray}
a two-way steering criterion is obtained
\begin{equation}
\sqrt{\langle\hat{A}_1\otimes \sigma_x\rangle^2+
\langle\hat{A}_2\otimes \sigma_z\rangle^2+\langle\hat{A}_3\otimes \sigma_y\rangle^2}\leqslant 1.\nonumber
\end{equation}

 For a $\mathcal{H}_d\otimes \mathcal{H}_d$ bipartite system, an operator is defined as
\begin{equation}
\label{MUBSH}
\hat{H}=\sum_{\mu=1}^2\sum_{a=0}^{d-1} \hat{\Pi}_{\mu}^a\otimes \vert\phi^a_{\mu}\rangle\langle\phi^a_{\mu}\vert,
\end{equation}
where $\{\vert\phi^a_{\mu}\rangle \}$ ($\mu=1,2$) are two sets of MUBs defined in Eq.~\eqref{MUB1}, while each $\{\hat{\Pi}_{\mu}^a\}$ is arbitrary. The one-way steering LSI from Alice to Bob, $\langle \hat{H}\rangle\leqslant 1+1/\sqrt{d}$,  appeared in previous works~\cite{Li,Zeng}.
As a generation, we
  introduce a one-parameter operator
\begin{eqnarray}
\hat{H}(\omega)&=& (1+\cos\omega)\left(\sum_a \hat{\Pi}_{1}^a\otimes \vert\phi^a_{1}\rangle\langle\phi^a_{1}\vert\right)\nonumber\\
\label{gmubs}
&+&(1-\cos\omega)\left(\sum_a \hat{\Pi}_{2}^a\otimes \vert\phi^a_{2}\rangle\langle\phi^a_{2}\vert\right).
\end{eqnarray}
Furthermore, we suppose that the two sets  of projective measurements, $\{\vert\phi^a_1\rangle\}$ and
 $\{\vert\phi^b_2\rangle\}$,  are related by a unitary
 transformation U with $U_{ab}$ as its matrix elements,
  $\vert \phi^a_1\rangle=U_{ab}\vert\phi^b_2\rangle+\sum_{c\neq b}U_{ac}\vert\phi^c_2\rangle$.
 
With the  calculation in Appendix~\ref{App1}, there is
\begin{equation}
\label{Eq31}
\beta^{\mathrm{A}\rightarrow \mathrm{B}}=1+\sqrt{\cos^2\omega+\sin^2\omega\vert U^{\mathrm{opt}}_{ab}\vert^2}.
\end{equation}
  where $\vert U^{\mathrm{opt}}_{ab}\vert$ has the  largest value among all the possible $\vert U_{ab}\vert, \forall a,b \in\{0,1,...,d-1\}$. If $\{\vert\phi^a_{\mu}\rangle \}$ ($\mu=1,2$) are   MUBs in Eq. \eqref{MUB1}, then $\vert U^{\mathrm{opt}}_{ab}\vert=1/\sqrt{d}$. By letting $\cos\omega=0$, the  one-way  LSI (from Alice to Bob), $\langle \hat{H}\rangle\leqslant 1+1/\sqrt{d}$, is recovered.

To derive the LSI from Bob to Alice, we can define
$\hat{\Pi}^a_{\mu}=\vert\psi^a_{\mu}\rangle\langle\psi^a_{\mu}\vert\ (\mu=1,2),\ \mathrm{with}\ \langle\psi^a_{\mu}\vert\psi^b_{\mu}\rangle=\delta_{ab}$,
and suppose that they are related   by a unitary operator $V$,
$\vert\psi^a_{1}\rangle=\sum_{b=0}^{d-1}V_{ab}\vert\psi^b_2\rangle$,
with $V_{ab}$ the matrix elements. Via a similar derivation, we shall get
\begin{equation}
\beta^{\mathrm{A}\leftarrow \mathrm{B}}_{\mathrm{ST}}=1+ \sqrt{\cos^2\omega+\sin^2\omega\vert V^{\mathrm{opt}}_{ab}\vert^2}.\nonumber
\end{equation}
Now, the steering threshold $\beta_{\mathrm{ST}}=\max\{\beta_{\mathrm{ST}}^{\mathrm{A}\rightarrow \mathrm{B}},\beta_{\mathrm{ST}}^{\mathrm{A}\leftarrow \mathrm{B}}\}$ is a function of $\omega$. For such cases, a function $\mathcal{R}(\omega)$ can be introduced and the two-way steering criterion can be rewritten in a standard form
\begin{equation}
\label{standard}
\mathcal{R}(\omega)\equiv\frac{\langle\hat{H}(\omega)\rangle}{\beta_{\mathrm{ST}}(\omega)}\leqslant 1.
\end{equation}
For a given bipartite state, one can first decide the expectation value $\langle\hat{H}(\omega)\rangle$ and then choose an optimal value for $\omega$, which makes $\mathcal{R}(\omega)$ have the largest value, through $d\mathcal{R}(\omega)/d\omega=0$.

After  demonstrating   that the two-way criteria can be constructed from the known one-way LSIs by introducing elaborately designed additional constraints and free parameters, we shall go back to the general operator $\hat{H}$ in Eq.~\eqref{defining-H}, where the numbers of the sets for $\{\hat{{\Pi}^a_\mu}\}$ and $\{\hat{M}^b_{\nu}\}$ are assumed to be finite, and show that the operator $\hat{H}$ can easily be generalized to the case where the experiment setting has a continuous form.

With a set of basis vectors, $\{\vert a \rangle\}_{a=0}^{d-1}$, a parameter $\omega$ can be used to label the experiment setting of Bob's measurements, $\hat{\Phi}^a_{\omega}={U}_{\omega}\vert a\rangle\langle a \vert {U}^{\dagger}_{\omega}$, where $U_{\omega}$ can take all the unitary operators in the $d$-dimensional unitary group $U(d)$. Consider the case that the probability for each measurement is equal-weighted, and the operator $\hat{H}$ can be designed as
\begin{equation}
\label{lsi5}
\hat{H}=\int d\mu_{\mathrm{Harr}}(\omega)\sum_{a=0}^{d-1}\hat{\Pi}_{\omega}^a\otimes \hat{\Phi}^a_{\omega},
\end{equation}
 where $d\mu_{\mathrm{Harr}}(\omega)$ is the Harr measure on the group $U(d)$. For the steering from Alice to Bob, by replacing each $\hat{\Pi}_{\omega}$ with $\mathfrak{p}(a\vert\omega,\xi)$,  there is
\begin{equation}
\label{Hatob}
\hat{H}^{\mathrm{A}\rightarrow \mathrm{B}}(\xi)=\int d\mu_{\mathrm{Harr}}(\omega)\sum_{a=0}^{d-1}\mathfrak{p}(a\vert\omega,\xi) \hat{\Phi}^a_{\omega}.\nonumber
\end{equation}
This type of operator was introduced in the recent work~\cite{WU2}, and based on the main results in Ref.~\cite{Wiseman1}, it was shown that
\begin{eqnarray}
\vert\hat{H}^{\mathrm{A}\rightarrow \mathrm{B}}(\xi)\vert^{\min}\geqslant \gamma_{\mathrm{ST}}^{\mathrm{A}\rightarrow \mathrm{B}}\equiv\frac{1}{d^2},\nonumber\\
\vert\hat{H}^{\mathrm{A}\rightarrow \mathrm{B}}(\xi)\vert^{\max}\leqslant\beta_{\mathrm{ST}}^{\mathrm{A}\rightarrow \mathrm{B}}\equiv\frac{H_d}{d},\nonumber
\end{eqnarray}
where $H_d=1+1/2+1/3+...+1/d$ is the Harmonic series. To construct the two-way LSI, some additional constraints for $\hat{\Pi}_{\omega}^a$ are required, and the following constraint is suggested
\begin{equation}
\label{constraint5}
\hat{\Pi}_{\omega}^a={U}^*_{\omega}\vert a\rangle\langle a \vert
{U}_{\omega}^{*\dagger},
\end{equation}
with ${U}^*_{\omega}$ to be the complex  conjugation of ${U}_{\omega}$. For   the steering from Bob  to Alice,  by replacing each $\hat{\Phi}_{\omega}$ with $\mathfrak{p}(a\vert\omega,\xi)$,  there is
 $\hat{H}^{\mathrm{A}\leftarrow \mathrm{B}}(\xi)=\int d\mu_{\mathrm{Harr}}(\omega)\sum_{a=0}^{d-1}\mathfrak{p}(a\vert\omega,\xi) \hat{\Pi}^a_{\omega}$. Obviously, $\hat{H}^{\mathrm{A}\leftarrow \mathrm{B}}(\xi)=[\hat{H}^{\mathrm{A}\rightarrow \mathrm{B}}(\xi)]^*$. Since a Hermitian operator and its complex conjugation have the same eigenvalues, the constraint above directly leads to $\gamma_{\mathrm{ST}}^{\mathrm{A}\leftarrow \mathrm{B}}=\gamma_{\mathrm{ST}}^{\mathrm{A}\rightarrow \mathrm{B}}$ and $\beta_{\mathrm{ST}}^{\mathrm{A}\leftarrow \mathrm{B}}=\beta_{\mathrm{ST}}^{\mathrm{A}\rightarrow \mathrm{B}}$. Based on these results, the two-way steering LSIs can be obtained
\begin{equation}
\label{lsi6}
\langle \hat{H}\rangle\geqslant \gamma_{\mathrm{ST}}\equiv\frac{1}{d^2},\ \langle \hat{H}\rangle \leqslant\beta_{\mathrm{ST}}\equiv\frac{H_d}{d}.
\end{equation}
Certainly, the above LSIs can also be derived with other types of constraints, say,
\begin{equation}
\label{constraint6}
\hat{\Pi}_{\omega}^a={U}_{\omega}\vert a\rangle\langle a \vert
{U}_{\omega}^{\dagger}.
\end{equation}

The Werner states can be defined as~\cite{Werner}
\begin{equation}
W^w_d=\frac{d-1+w}{d-1}\frac{I_d \otimes I_d}{d^2}-\frac{w}{d-1}\frac{\mathbf{V}}{d},\nonumber
\end{equation}
where $0\leqslant w\leqslant 1$ and $\mathbf{V}$ is the ``flip" operator defined by $\mathbf{V}\vert\phi\rangle\otimes\vert\psi\rangle=\vert\psi\rangle\otimes\vert\phi\rangle$. With the constraint in Eq.~\eqref{constraint6}, there is $\langle\hat{H}\rangle\equiv \mathrm{Tr}(\hat{H}W^w_d)=(1-w)/d$. With the first LSI in Eq.~\eqref{lsi6}, the Werner state is verified to be two-way steerable if $1-w< 1/d$. The steering threshold, $\gamma_{\mathrm{ST}}\equiv1/d^2$, is a tight bound since the known fact that Werner is non-steerable iff ${1-w}\geqslant 1/d$~\cite{Wiseman1}.

For a mixing  parameter $\eta$, the $d$-dimensional isotopic state is defined as
\begin{equation}
W^{\eta}_d=(1-\eta)\frac{I_d \otimes I_d}{d^2}+\eta \mathbf{P}_+,\nonumber
\end{equation}
where $\mathbf{P}_+=\vert\psi_+\rangle\langle\psi_+\vert$, and $\vert\psi_+\rangle=\sum_{i=1}^d\vert i\rangle\vert i\rangle/\sqrt{d}$ is a maximally entangled state. With the constraint in Eq.~\eqref{constraint5}, there is
$\langle\hat{H}\rangle\equiv\mathrm{Tr}(\hat{H}W^{\eta}_d)=[1+(d-1)\eta]/d$. With the second LSI in Eq.~\eqref{lsi6}, the isotopic state is shown to be two-way steerable if $1+(d-1)\eta>H_d$.  It is known that the isotopic state is unsteerable iff $1+(d-1)\eta\leqslant H_d$~\cite{Wiseman1}, the steering threshold, $\beta_{\mathrm{ST}}\equiv H_d/d$, is also a tight bound.

\subsection{Accompanied linear steering inequality}
Besides designing the two-way criteria from the known one-way LSIs, there are other methods, and in the following, how to construct two-way LSIs from the Bell operators will be discussed. Formally, a Bell inequality is expressed as $\langle \hat{\mathcal{B}}\rangle\equiv\mathrm{Tr}(\hat{\mathcal{B}}\rho_{AB})\leqslant\beta_{\mathrm{NL}}$, with $\beta_{\mathrm{NL}}$ the nonlocal boundary. Following the discussion above, one can first derive the two one-way LSIs, $\langle \hat{\mathcal{B}}\rangle\leqslant\beta_{\mathrm{ST}}^{\mathrm{A}\rightarrow \mathrm{B}}$ and $\langle \hat{\mathcal{B}}\rangle\leqslant\beta_{\mathrm{ST}}^{\mathrm{A}\leftarrow \mathrm{B}}$, and then, from the definition, $\beta_{\mathrm{ST}}=\max\{\beta_{\mathrm{ST}}^{\mathrm{A}\rightarrow\mathrm{B}},\beta_{\mathrm{ST}}^{\mathrm{A}\leftarrow\mathrm{B}}\}$, the two-way inequality can be expressed as $\langle\hat{\mathcal{B}}\rangle\leqslant\beta_{\mathrm{ST}}$. In this work, these types of LSIs are referred as the accompanied linear steering inequalities (ALSIs). From the  theory of steering, the nonlocality is more stronger than steering, and in general, there is a simple relation between $\beta_{\mathrm{ST}}$ and  $\beta_{\mathrm{NL}}$: $\beta_{\mathrm{ST}}\leqslant \beta_{\mathrm{NL}}$.

Firstly, let us derive the LSIs accompanied with the CHSH inequality~\cite{chsh}. Let $\hat{A}_1$,$\hat{A}_2$, $\hat{B}_1$ and $\hat{B}_2$ be the $2$-value operators, and with the conventional denotation, $\langle AB\rangle=\langle \hat{A}\otimes \hat{B}\rangle$, the CHSH inequality can be expressed as
\begin{equation}
\label{chsh}
\langle A_1B_1\rangle +\langle A_1B_2\rangle+\langle A_2B_1\rangle-\langle A_2B_2\rangle
\leqslant\beta_{\mathrm{NL}}=2.\nonumber
\end{equation}
Let us consider the steering from Alice to Bob, and with $\hat{\mathcal{B}}=\hat{A}_1\otimes(\hat{B}_1+\hat{B}_2)+\hat{A}_2\otimes(\hat{B}_1-\hat{B}_2)$ and Eq.~\eqref{ddddd}, one can have four possible $\hat{H}(\xi)$,
\begin{equation}
\hat{H}_{++}=2\hat{B}_1, \hat{H}_{+-}=2\hat{B}_2, \hat{H}_{-+}=-2\hat{B}_2,\hat{H}_{--}=-2\hat{B}_1.\nonumber
\end{equation}
The one-way steering threshold can be obtained $\beta^{\mathrm{A}\rightarrow \mathrm{B}} _{\mathrm{ST}}=\max\{\vert \hat{H}_{\pm\pm}\vert^{\max}\}$, and certainly, $\beta^{\mathrm{A}\rightarrow \mathrm{B}}_{\mathrm{ST}}=2$. Now, with two arbitrary two-dimensional unitary operators $U$ and $V$, and introducing following additional constraints
\begin{equation}
\label{constraint8}
\hat{B}_1=U\sigma_xU^{\dagger}, \ \hat{B}_2=V\sigma_zV^{\dagger},
\end{equation}
it can be found that $\beta^{\mathrm{A}\rightarrow \mathrm{B}}_{\mathrm{ST}}$ remains unchanged with the constraints above. This property of the CHSH operator is very different from the ones  discussed above.

For the steering from Bob to Alice, another one-way steering threshold, which can be obtained with the same method, is $\beta^{\mathrm{A}\leftarrow \mathrm{B}}_{\mathrm{ST}}=2$.  With two arbitrary two-dimensional unitary operator $\bar{U}$ and $\bar{V}$ and the following additional constraints,
\begin{equation}
\label{constraint9}
\hat{A}_1=\bar{U}\sigma_x\bar{U}^{\dagger},\ \hat{A}_2=\bar{V}\sigma_z\bar{V}^{\dagger},
\end{equation}
one can easily verify that $\beta^{\mathrm{A}\leftarrow \mathrm{B}}_{\mathrm{ST}}$ keeps unchanged with these constraints. From the definition in Eq.~\eqref{beta},  ALSI is known as
\begin{equation}
 \label{achsh}
\langle A_1B_1\rangle +\langle A_1B_2\rangle+\langle A_2B_1\rangle-\langle A_2B_2\rangle
\leqslant\beta_{\mathrm{ST}}=2.
\end{equation}

In the present work, the steering threshold is said to be LHS-attainable if it can be attained by the compatible measurement performed either by Alice or Bob. For the maximally entangled state $\vert\Phi\rangle=(\vert 00\rangle+\vert11\rangle)/\sqrt{2}$, Alice can perform the compatible measurement: $\hat{\Pi}^a_1=(I_2+(-1)^a\sigma_x/\sqrt{2})/2$, and $\hat{\Pi}^b_2=(I_2+(-1)^b\sigma_z/\sqrt{2})/2$. The measurements performed by Bob are chosen as $\hat{B}_1=(\sigma_x+\sigma_z)/\sqrt{2}$ and $\hat{B}_2=(\sigma_x-\sigma_z)/\sqrt{2}$. Under such conditions, there is $\langle \hat{\mathcal{B}}\rangle=2=\beta_{\mathrm{ST}}$.  Therefore, $\beta_{\mathrm{ST}}=2$ is LHS-attainable.

Now, let us return to the operator in Eq.~\eqref{defining-H}, it can be shown that the experiment setting for Alice may be different from the one for Bob. An example for such situations is in below. In the Pironio inequality~\cite{Pironio}, the measurements performed by Alice are  two-result measurements satisfying
\begin{equation}
\label{Alice}
\sum_{a=0}^1 \hat{\Pi}^a_\mu=I_d,\ \mu\in\{0,1,...,d-1\},
\end{equation}
with $d$ sets of $\{\hat{\Pi}_{\mu}^a\}$ for a $d$-dimensional system, while the measurements performed by Bob are fixed as
\begin{equation}
\label{Bob}
\sum_{b=0}^{d-1}\hat{M}^b_0=I_d,\  \sum_{b=0}^{1}\hat{M}^{b}_1=I_d,
\end{equation}
with just two sets of $\{\hat{M}_{\nu}^b\}$.
The Pironio inequality can be applied for disproving the Peres conjecture by showing Bell nonlocality from bound entanglement \cite{ vert,You,Pal}.

With the denotations  $p(ab\vert \mu\nu)=\langle \hat{\Pi}^a_{\mu}\otimes \hat{M}^b_{\nu}\rangle$, $p_A(a\vert \mu)=\langle \hat{\Pi}^a_{\mu}\otimes I_B\rangle$, and $p_B(b\vert \nu) =\langle I_A\otimes \hat{M}^b_{\nu}\rangle$, the Pironio inequality for $d=3$,  which was used in Ref.~\cite{vert}, takes the form
\begin{eqnarray}
&-&p_A(0\vert 2)-2p_B(0\vert 1)-p(01\vert00)-p(00\vert 10)+p(00\vert20) \nonumber\\
&+&p(01\vert 20)+p(00\vert 01)+p(00\vert11)+p(00\vert 21)\leqslant \beta_{\mathrm{NL}}=0.\nonumber
\end{eqnarray}
To derive the Bell operator, the term $2 p_B(0\vert 1)$ is treated with an equivalent form:  $2 p_B(0\vert 1)=\langle(\hat{\Pi}^0_0+\hat{\Pi}^1_0)\otimes \hat{M}^0_1\rangle+\langle(\hat{\Pi}^0_1+\hat{\Pi}^1_1)\otimes \hat{M}^0_1\rangle$, and using Eqs.~\eqref{Alice} and~\eqref{Bob}, the Bell operator for above inequality can be constructed,
\begin{eqnarray}
\hat{\mathcal{B}}_3&=&-\hat{{\Pi}}^0_0\otimes\hat{M}^1_0-\hat{{\Pi}}^1_0\otimes\hat{M}^0_1-\hat{{\Pi}}^0_1\otimes\hat{M}^0_0\nonumber\\
&&-\hat{{\Pi}}^1_1\otimes\hat{M}^0_1+\hat{\Pi}^0_2\otimes (\hat{M}_1^0-\hat{M}^2_0).\nonumber
\end{eqnarray}
For the steering from Alice to Bob,  the operators $\hat{H}_{k_1k_2k_3}$ are listed below:
\begin{eqnarray}
\hat{H}_{000}&=&-\hat{M}^1_1,\ \ \ \ \ \ \ \ \ \ \ \ \hat{H}_{001}=-\hat{M}^1_0-\hat{M}^0_0,\nonumber\\
 \hat{H}_{010}&=&-\hat{M}^1_0-\hat{M}_0^2, \ \hat{H}_{011}=-\hat{M}^0_1-\hat{M}^1_0, \nonumber\\
\hat{H}_{100}&=&-\hat{M}^0_0-\hat{M}_0^2,\  \hat{H}_{101}=-\hat{M}^0_0-\hat{M}^0_1,\nonumber\\
 \hat{H}_{110}&=&-\hat{M}^0_1-\hat{M}_0^2,\ \hat{H}_{111}=-2\hat{M}^0_1.
\end{eqnarray}
From the elementary property of POVM, $\langle\phi\vert \hat{M}^b_y\vert\phi\ket\geqslant0$, one may have $\vert \hat{H}_{k_1k_2k_3}\vert^{\max}\leqslant 0$.

For the steering from  Bob to  Alice, the  $\hat{H}_{k_1k_2}$ are known as
\begin{eqnarray}
\hat{H}_{00}&=&-\hat{\Pi}^0_1-\hat{\Pi}^1_2, \hat{H}_{01}=-\hat{\Pi}^0_1, \nonumber\\
\hat{H}_{10}&=&-\hat{\Pi}^0_0-\hat{\Pi}^1_2, \hat{H}_{11}=-\hat{\Pi}^0_0, \nonumber\\
\hat{H}_{20}&=&-\hat{\Pi}^0_2-\hat{\Pi}^1_2, \hat{H}_{21}=-\hat{\Pi}^0_2, \nonumber
\end{eqnarray}
and with the result  $\vert\hat{H}_{k_1k_2}\vert^{\max}\leqslant0$, there is $\beta^{\mathrm{A}\leftarrow \mathrm{B}}_{\mathrm{ST}}=0$. By jointing it with the result $\beta^{\mathrm{A}\rightarrow \mathrm{B}}_{\mathrm{ST}}=0$, we have $\beta_{\mathrm{ST}}=0$.

For the general $\mathcal{H}_d\otimes \mathcal{H}_d$ case, one may define the Bell operator
\begin{eqnarray}
\hat{\mathcal{B}}_d&=&\hat{\Pi}^0_0 \otimes (\hat{M}^0_1-\hat{M}^0_0)\nonumber\\
&-&\left[\sum_{i=1}^{d-1}\left(\hat{\Pi}^0_i\otimes \hat{M}^i_0+\hat{\Pi}^1_i\otimes \hat{M}^0_1\right)\right].\nonumber
\end{eqnarray}
For the steering from Alice to Bob, besides $\hat{H}_{00\cdots0}=-\hat{M}^1_1$ and $\hat{H}_{11\cdots1}=-(d-1)\hat{M}^0_1$, the operators $\hat{H}_{k_1k_2\cdots k_d}$ can be expressed as
$\hat{H}_{k_1k_2\cdots k_d}=-(c_0 \hat{M}^0_1+ \sum_{j=1}^d c_j \hat{M}^j_0)$, with $c_j\in\{0,1,2,...,d-1\}$ and $\sum_{j=0}^{d-1}c_j=d-1$. Similarly with the derivation for the case $d=3$, one can obtain $\beta^{\mathrm{A}\rightarrow \mathrm{B}}_{\mathrm{ST}}=0$.

For the steering from  Bob to  Alice, there are
\begin{eqnarray}
\hat{H}_{00}&=&-\sum_{j=1}^{d-1}\hat{\Pi}^1_{j},\ \ \ \ \ \ \ \ \ \ \ \  \hat{H}_{01}=-\hat{\Pi}^0_0,\nonumber\\
\hat{H}_{i0}&=&\hat{\Pi}^0_0-(d-1)I_d,\ \hat{H}_{i1}=-\hat{\Pi}^0_i,\nonumber
\end{eqnarray}
with $i\in\{1, 2,..., d-1\}$. Obviously, $\beta^{\mathrm{A}\leftarrow \mathrm{B}}_{\mathrm{ST}}=0$, and finally, one can have an LSI, $\langle \hat{\mathcal{B}}_d\rangle\leqslant\beta_{\mathrm{ST}}=0$, which is accompanied to  the Pironio equality in following version~\cite{You}:
\begin{equation}
p(00\vert01)-p(00\vert00)-\sum_{i=1}^{d-1}\left[p(0i\vert i0)+p(10\vert i1)\right]\leqslant\beta_{\mathrm{NL}}=0.\nonumber
\end{equation}

For the Bell operator
\begin{equation}
\label{tilted}
\hat{\mathcal{B}}=\delta\hat{A}_1\otimes I_{B}+\alpha \hat{A}_1\otimes (\hat{B}_1+\hat{B}_2)+\hat{A}_2\otimes (\hat{B}_1-\hat{B}_2),
\end{equation}
where $\delta\geqslant0$, $\alpha\geqslant1$, and $\hat{A}_{\mu}$ and $\hat{B}_{\nu}$ are arbitrary 2-value operators, the so-called tilted CHSH inequality, $\langle\hat{\mathcal{B}}\rangle\leqslant\beta_{\mathrm{NL}}\equiv\delta +2\alpha$, has been introduced~\cite{Massar}. Now, we first consider the one-way ALSI from Bob to Alice, and by replacing each $\hat{B}_{\nu}$ with $\mathfrak{d}(\nu\vert\xi)$ defined in Eq.~\eqref{ddddd}, one can have four possible $\hat{H}(\xi)$,
\begin{eqnarray}
\hat{H}_{++}&=&(\delta+2\alpha)\hat{A}_1,\ \ \hat{H}_{+-}=\delta\hat{A}_1+2\hat{A}_2,\nonumber\\
\hat{H}_{-+}&=&\delta\hat{A}_1-2\alpha\hat{A}_2,\hat{H}_{--}=(\delta-2\alpha)\hat{A}_1.\nonumber
\end{eqnarray}
 With a simple calculation, there is $\vert\hat{H}_{\pm\pm}\vert\leqslant\beta^{\mathrm{A}\leftarrow \mathrm{B}}_{\mathrm{ST}}\equiv\delta+2\alpha$, and this steering threshold remains unchanged if the constraints in Eq. \eqref{constraint9} are introduced.

For the steering from Alice to Bob, by replacing each $\hat{A}_{\mu}$ with its corresponding $\mathfrak{d}(\mu\vert\xi)$, there are,
\begin{eqnarray}
\hat{H}_{++}&=&\delta I_B+(\alpha+1)\hat{B}_1+(\alpha-1)\hat{B}_2, \hat{H}_{--}=-\hat{H}_{++},\nonumber\\
\hat{H}_{+-}&=&\delta {I}_B +(\alpha-1)\hat{B}_1+(\alpha+1)\hat{B}_2, \hat{H}_{-+}=-\hat{H}_{+-}.\nonumber
\end{eqnarray}
If $\hat{B}_1$ and $\hat{B}_2$ are arbitrary, one can have $\vert\hat{H}_{\pm\pm}\vert\leqslant\beta^{\mathrm{A}\rightarrow
\mathrm{B}}_{\mathrm{ST}}\equiv\delta+2\alpha$. However,
when the additional conditions in Eq. \eqref{constraint8}  are introduced, it can be found that $\beta^{\mathrm{A}\rightarrow \mathrm{B}}_{\mathrm{ST}}$ dose not remain unchanged. For example, with the suitable chosen unitary operators $U$ and $V$, one may have a special constraint
\begin{equation}
\label{constraint10}
\hat{B}_1+\hat{B}_2=\sqrt{2}\sigma_z, \hat{B}_1-\hat{B}_2=\sqrt{2}\sigma_x,
\end{equation}
and under this condition, there should be $\beta^{\mathrm{A}\rightarrow \mathrm{B}}_{\mathrm{ST}}=\delta+\sqrt{2(\alpha^2+1)}$. With the constraint $\alpha\geqslant1$ in Eq. \eqref{tilted}, one can find that: (a) If $\alpha=1$, $\beta^{\mathrm{A}\rightarrow \mathrm{B}}_{\mathrm{ST}}= \beta_{\mathrm{NL}}$ and (b) If $\alpha>1$, $\beta^{\mathrm{A}\rightarrow \mathrm{B}}_{\mathrm{ST}}$    is always lower than the nonlocal threshold $\beta_{\mathrm{NL}}\equiv\delta+2\alpha$.  Now, define a new parameter $\bar{\delta}=\sqrt{2}\delta/2$, and based on the denotations above, the one-way LSI from Alice to Bob, $\langle\hat{\mathcal{B}}\rangle\leqslant \delta+\sqrt{2(\alpha^2+1)}$, can be expressed with an equivalent form
\begin{equation}
\bar{\delta}\langle A_1\rangle+\alpha\langle A_1\sigma_z\rangle+\langle A_2\sigma_x\rangle\leqslant \bar{\delta}+\sqrt{\alpha^2+1}.
\end{equation}
The above inequality has already  been derived in Ref.~\cite{kwek} with a different method. This criterion has been applied for the task of robust semi-device-independent certification.

In the end of this section, a simple example will be given to show that the two-way steering LSI can also be derived from the entanglement witness. The entanglement witness $\mathcal{W}$ is an operator satisfying the following two conditions: (1) For any separable state $\rho_{\mathrm{AB}}^{\mathrm{sep}}$, $\mathrm{Tr}(\mathcal{W}\rho_{\mathrm{AB}}^{\mathrm{sep}})\geqslant0 $ and (2) There exists at least one entangled state $\rho_{\mathrm{AB}}^{\mathrm{ent}}$ such that $\mathrm{Tr}(\mathcal{W}\rho_{\mathrm{AB}}^{\mathrm{ent}})< 0$ \cite{Guhne}. Consider a simple entanglement witness
\begin{equation}
\mathcal{W}=\frac{1}{2}I_2\otimes I_2-\vert\psi_+\rangle\langle\psi_+\vert,\nonumber
\end{equation}
where $\vert\psi_+\rangle=(\vert00\rangle+\vert11\rangle)/\sqrt{2}$, and it can be rewritten as
\begin{equation}
\mathcal{W}=\frac{1}{4}(I_2\otimes I_2-\sigma_x\otimes \sigma_x+\sigma_y\otimes \sigma_y-\sigma_z\otimes\sigma_z).\nonumber
\end{equation}
With a simple calculation, one can have the two-way LSI,
\begin{equation}
\frac {1-\sqrt{3}}{4}\leqslant\langle \mathcal{W}\rangle\leqslant\frac{1+\sqrt{3}}{4},\nonumber
\end{equation}
which is designed from the entanglement witness.

\section{Detecting genuine multipartite  two-way steerability}
\label{Sec4}
As shown in introduction, a fundamental property that steering is inherently asymmetric with respect to the observers~\cite{bowles,Midgley} is quite different from the quantum nonlocality and entanglement. In this section, the concept of genuine two-way steerability for multipartite system will be introduced first, and then it will be shown that the genuine two-way steerability for multipartite system can be verified by the LSIs.

\subsection{Genuine multipartite two-way steerability}
 For a three-particle system $\rho_{ABC}$, the state is fully separable if it can be written as $\rho^\mathrm{fs}_{ABC}=\sum_k p_k\rho^A_k\otimes \rho^B_k\otimes \rho_k^C$, where $p_k$ form a probability distribution. If a state is not of this form, it is entangled. A state is biseparable if it takes the form like $\rho_{AB:C}^\mathrm{bs}=\sum_k p_k \rho^{AB}_k\otimes \rho^{C}_k$. More generally, mixture of bisparable states for different partitions are still biseparable~\cite{Guhne,Horos},
\begin{equation}
\label{gme}
\rho^\mathrm{bs}_{ABC}=p_1\rho^\mathrm{bs}_{AB:C}
+p_2\rho^\mathrm{bs}_{AC:B}+p_3\rho^\mathrm{bs}_{BC:A},
\end{equation}
and a state, which is  not biseparable,  is  genuine multipartite entangled (GME).

The first definition of genuine multipartite nonlocality was proposed by Svetlichny~\cite{svet}. For a Hilbert space $\mathcal{H}_{A}\otimes\mathcal{H}_{B}\otimes\mathcal{H}_C$, POVMs can be defined for each local Hilbert space: $\sum_a\hat{\Pi}_x^a=I_A,\sum_b\hat{M}_y^b=I_B,\sum_c\hat{N}_z^c=I_C$, and the expectation $\langle\hat{\Pi}_x^a\otimes \hat{M}_y^b\otimes\hat{N}_z^c\rangle$ is denoted by $p(abc\vert xyz)=\langle\hat{\Pi}_x^a\otimes \hat{M}_y^b\otimes\hat{N}_z^c\rangle$. For an LHV model, $p(abc\vert xyz)$ can be written in the form~\cite{svet}
\begin{eqnarray}
\label{gmnl}
p(abc\vert xyz)&=&\int d\lambda q(\lambda)p_{\lambda}(ab\vert xy)p_{\lambda}(c\vert z)\nonumber\\
&+&\int d\mu q(\mu)p_{\mu}(ac\vert xz)p_{\mu}(b\vert y)\nonumber\\
&+&\int d\nu q(\nu)p_{\nu}(bc\vert yz)p_{\nu}(a\vert x),
\end{eqnarray}
where $\int d\lambda q(\lambda)+\int d\mu q(\mu)+\int d\nu q(\nu)=1$. The convex combination represents a situation where only two parties share a nonlocal resource in any measurement run. On the other hand, if $p(abc\vert xyz)$ for a given state $\rho$ cannot be written in the above form, it is necessary that the three parties must share some common nonlocal resource, and then the state is  genuine multipartite nonlocal (GMNL). Now, it is possible to write down Bell inequalities, $\langle \hat{\mathcal{B}}\rangle\leqslant\beta_{\mathrm{GMNL}}$, with $\beta_{\mathrm{GMNL}}$ the threshold for multipartite nonlocality. If the inequality is violated, one may conclude that the correlations  $p(abc\vert xyz)$ are genuine multipartite nonlocal.

 Following the general theory in Ref.~\cite{Wiseman1}, we shall give a definition for the genuine multipartite steerability  (GMST) with the following   two constraints: (1) The genuine multipartite entanglement is weaker than the genuine multipartite steerability,
\begin{equation}
\label{st>e}
\forall \rho_{\mathrm{GMST}}\in\{\rho_{\mathrm{GME}}\},
\end{equation}
which declares that  each $ \rho_{\mathrm{GMST}}$ belongs to the set of GME states; and (2) The genuine multipartite nonlocality is stronger than the genuine multipartite steerability
\begin{equation}
\label{nl>st}
\forall \rho_{\mathrm{GMNL}}\in\{\rho_{\mathrm{GMST}}\},
\end{equation}
and every GMNL state must be a GMST state.  Under these two conditions, a multipartite LHS model, which is based on the fundamental definition of the LHS model in Eq.~\eqref{tilderho}, can be constructed.

For a given partition $AB:C$, according to  Eq.~\eqref{tilderho}, one can easily have the definitions for the three-particle state to be steerable from AB to C ($AB\rightarrow C$) and C to AB ($C\rightarrow AB$) through the following four steps: (1) When the set of measurements $\{\hat{\Pi}_x^a\otimes \hat{M}_y^b\otimes\hat{N}_z^c\}$ is performed, the conditional states are defined as $\tilde{\rho}^{ab}_{xy}=\mathrm{Tr}(\rho_{ABC} \hat{\Pi}_x^a\otimes \hat{M}_y^b\otimes I_C)$ and $\tilde{\rho}^{c}_{z}=\mathrm{Tr}(\rho_{ABC} I_A\otimes I_B\otimes \hat{N}^c_z)$. (2) If the assemblage $\{\tilde{\rho}^{ab}_{xy}\}$ does not admit an LHS model like
\begin{equation}
\tilde{\rho}^{ab}_{xy}=\int d\xi_1\Omega(\xi_1)\mathfrak{p}(ab\vert xy,\xi_1)\rho_{\xi_1},
\end{equation}
we say the state is steerable from AB to C. An equivalent version of the above equation is
\begin{equation}
\label{abc}
p(abc\vert xyz)=\int d\xi_1\Omega(\xi_1)\mathfrak{p}(ab\vert xy,\xi_1)\mathrm{Tr}(\hat{N}^c_z\rho_{\xi_1}).
\end{equation}
(3) If the assemblage $\{\tilde{\rho}^{c}_{z}\}$ cannot be expanded as
 \begin{equation}
\tilde{\rho}^{c}_{z}=\int d\xi_2\Omega(\xi_2)\mathfrak{p}(c\vert z,\xi_2)\rho_{\xi_2},
\end{equation}
the three-particle state is steerable from C to AB. The equation above can also be expressed as
\begin{equation}
\label{cab}
p(abc\vert xyz)=\int d\xi_2\Omega(\xi_2)\mathfrak{p}(c\vert z,\xi_2)\mathrm{Tr}(\hat{\Pi}_x^{a}\otimes \hat{M}^b_y \rho_{\xi_2}).
\end{equation}
(4) By jointing Eq.~\eqref{abc} and Eq.~\eqref{cab} together, for the given partition $AB:C$, one can define the two-way steering: If the correlation $p(abc\vert xyz)$ cannot be  expanded as
\begin{eqnarray}
\label{abcba}
p(abc\vert xyz)&=&q_1\int d\xi_1\Omega(\xi_1)\mathfrak{p}(ab\vert xy,\xi_1)\mathrm{Tr}(\hat{N}^c_z\rho_{\xi_1})\nonumber\\
&+&q_2\int d\xi_2\Omega(\xi_2)\mathfrak{p}(c\vert  z,\xi_2)\mathrm{Tr}(\hat{\Pi}^{a}_x\otimes \hat{M}^b_y \rho_{\xi_2}),\nonumber\\
\end{eqnarray}
with $q_1+q_2=1$, the three-particle state is two-way steerable, $AB\leftrightharpoons C$. The definition for $AC\leftrightharpoons B$ and $BC\leftrightharpoons A$ can be constructed similarly. By collecting all the definitions together, a generalized LHS model can be obtained for the correlation $p(abc\vert xyz)$:
\begin{eqnarray}
\label{abccba}
p(abc\vert xyz)&=&q_1\int d\xi_1\Omega(\xi_1)\mathfrak{p}(ab\vert xy,\xi_1)\mathrm{Tr}(\hat{N}^c_z\rho_{\xi_1})\nonumber\\
&+&q_2\int d\xi_2\Omega(\xi_2)\mathfrak{p}(c\vert  z,\xi_2)\mathrm{Tr}(\hat{\Pi}^{a}_x\otimes \hat{M}^b_y \rho_{\xi_2})\nonumber\\
&+&q_3\int d\xi_3\Omega(\xi_3)\mathfrak{p}(ac\vert xz,\xi_3)\mathrm{Tr}(\hat{M}^b_y\rho_{\xi_3})\nonumber\\
&+&q_4\int d\xi_4\Omega(\xi_4)\mathfrak{p}(b\vert  y,\xi_4)\mathrm{Tr}(\hat{\Pi}^{a}_x\otimes \hat{N}^c_z \rho_{\xi_4})\nonumber\\
&+&q_5\int d\xi_5\Omega(\xi_5)\mathfrak{p}(bc\vert yz,\xi_5)\mathrm{Tr}(\hat{\Pi}^a_x\rho_{\xi_5})\nonumber\\
&+&q_6\int d\xi_6\Omega(\xi_6)\mathfrak{p}(a\vert  x,\xi_6)\mathrm{Tr}(\hat{M}^{b}_y\otimes \hat{N}^c_z \rho_{\xi_6}),\nonumber\\
\end{eqnarray}
where $\int d\xi_k\Omega(\xi_k)=1$ and $\sum_{k=1}^6q_k=1$. If  $p(abc\vert xyz)$ do not admit such a model, then  the state is called  genuine multipartite (two-way) steerable. The above definition can straightforwardly be extended to the case with more than three particles.

It can be shown that the constraint in Eq.~\eqref{st>e} is satisfied: All the biseparable states in Eq.~\eqref{gme} always admits the generalized LHS models. First, let $\rho^\mathrm{bs}_{AB:C}=\int d \xi \Omega(\xi) \rho^{AB}(\xi)\otimes \rho^C(\xi)$, which is a usual definition for the biseparable state, and then
\begin{equation}
\label{case1}
p(abc\vert xyz)=\int d \xi \Omega(\xi) \mathrm{Tr}\left[\hat{\Pi}^a_x\otimes \hat{M}^b_y\rho^{AB}(\xi)\right]\mathrm{Tr}\left[\hat{N}^c_z\rho^{c}(\xi)\right].
\end{equation}
Let $\mathfrak{p}(ab\vert xy,\xi)=\mathrm{Tr}[\hat{\Pi}^a_x\otimes \hat{M}^b_y\rho^{AB}(\xi)]$, and $p(abc\vert xyz)$ in Eq.~\eqref{case1} is similar with the first term in Eq.~\eqref{abccba}. Second, let $\mathrm{Tr}[\hat{N}^c_z\rho^{c}(\xi)]=\mathfrak{p}(c\vert  z,\xi)$, and $p(abc\vert xyz)$ in Eq.~\eqref{case1} is similar with the second term in Eq.~\eqref{abccba}. For the same reason, one may verify that $p(abc\vert xyz)$ for $\rho^\mathrm{bs}_{AC:B}$ can be expressed by the third and fourth terms in Eq.~\eqref{abccba}, and $p(abc\vert xyz)$ for $\rho^{bs}_{BC:A}$ can be expressed with the last two terms in Eq.~\eqref{abccba}.

The first term on the right hand of Eq.~\eqref{abccba} can be rewritten as $q_1\int d\xi_1\Omega(\xi_1)\mathfrak{p}(ab\vert xy,\xi_1)\mathfrak{p}(c\vert z, \xi_1)$ with $\mathfrak{p}(c\vert z,\xi_1)=\mathrm{Tr}(\hat{N}^c_z\rho_{\xi_1})$, and the second term can be expressed as $q_2\int d\xi_2\Omega(\xi_2)\mathfrak{p}(c\vert  z,\xi_2)\mathfrak{p}(ab \vert xy, \xi_2)$ with $\mathfrak{p}(ab \vert xy, \xi_2)=\mathrm{Tr}(\hat{\Pi}^{a}_x\otimes \hat{M}^b_y \rho_{\xi_2})$. Both the integrals belong to the first term on the right hand of Eq.~\eqref{gmnl}. Furthermore, one may verify that the third and fourth term on the right hand of Eq.~\eqref{abccba} belong to the second term on the right hand of Eq.~\eqref{gmnl}, and the last two terms in Eq.~\eqref{abccba} belong to the final term in Eq.~\eqref{gmnl}. Based on these results, one may conclude that the constraint in Eq.~\eqref{nl>st} does hold since the generalized LHS model in Eq.~\eqref{abccba} belongs to the general LHV model in Eq.~\eqref{gmnl}.

\subsection{LSIs for multipartite system}

The LSI in Sec.~\ref{Sec3} is designed for detecting the steerability for bipartite system, and to construct LSI for multipartite system, some denotations should first be introduced. Instead of the usually used symbol $\rho_{ABC}$,  $\rho_{123}$ is used to denote  the three-partite state. Furthermore, the symbol $ij\rightarrow k$ , with $i\neq j\neq k$, and $i,j,k\in\{1,2,3\}$, represents the case that the particles $i$ and $j$ are on Alice's side, and the particle $k$ is on Bob's side. At the same time, it is always supposed that only the measurements on Bob's side are trusted. For a Hermitian operator $\hat{H}$,
\begin{equation}
\label{hermitian}
\hat{H}=\sum_{a,b,c}\sum_{x,y,z}\gamma(abc\vert xyz)\hat{\Pi}^a_x\otimes \hat{M}^b_y\otimes \hat{N}^c_z,
\end{equation}
the operator $\hat{H}^{12\rightarrow 3}$ can be defined
\begin{equation}
\label{123}
\hat{H}^{12\rightarrow 3}=\sum_{a,b,c}\sum_{x,y,z}\gamma(abc\vert xyz)\mathfrak{p}(ab\vert xy,\xi)\hat{N}^c_z,
\end{equation}
and it can be understood that the operator $\hat{\Pi}^a_x\otimes \hat{M}^b_y$ in Eq.~\eqref{hermitian} is replaced by $\mathfrak{p}(ab\vert xy,\xi)$, which is the predetermined value of $\hat{\Pi}^a_x\otimes \hat{M}^b_y$ in an LHV model. Similarly, the operator $\hat{H}^{1\rightarrow 23}$ is
\begin{equation}
\label{1123}
\hat{H}^{1\rightarrow 23}=\sum_{a,b,c}\sum_{x,y,z}\gamma(abc\vert xyz) \mathfrak{p}(a\vert x,\xi)\hat{M}^b_y\otimes \hat{N}^c_z.
\end{equation}
The rest ones can be constructed in a similar way.

Now, suppose  that  correlation $p(abc\vert xyz)$ for $\rho_{123}$ admits the generalized LHS model in Eq.~\eqref{abccba}, and one may obtain
\begin{eqnarray}
\langle\hat{H}\rangle &=&q_1\int d\xi_1\Omega(\xi_1)\mathrm{Tr}(\hat{H}^{12\rightarrow 3} \rho_{\xi_1})\nonumber\\
&+&q_2\int d\xi_2\Omega(\xi_2)\mathrm{Tr}(\hat{H}^{3\rightarrow 12} \rho_{\xi_2})\nonumber\\
&+&q_3\int d\xi_3\Omega(\xi_3)\mathrm{Tr}(\hat{H}^{13\rightarrow 2}\rho_{\xi_3})\nonumber\\
&+&q_4\int d\xi_4\Omega(\xi_4)\mathrm{Tr}(\hat{H}^{2\rightarrow 13} \rho_{\xi_4})\nonumber\\
&+&q_5\int d\xi_5\Omega(\xi_5)\mathrm{Tr}(\hat{H}^{23\rightarrow 1}\rho_{\xi_5})\nonumber\\
&+&q_6\int d\xi_6\Omega(\xi_6)\mathrm{Tr}(\hat{H}^{1\rightarrow 23} \rho_{\xi_6}).
\end{eqnarray}
The threshold for genuine multipartite two-way steerability $\beta_{\mathrm{GMST}}$ is defined as
\begin{equation}
\label{betamax}
\beta_{\mathrm{GMST}}=\max_{ijk}\{\vert\hat{H}^{ij\rightarrow k}\vert^{\max},\vert\hat{H}^{k\rightarrow ij}\vert^{\max}\},
\end{equation}
where $i\neq j\neq k$ and $i,j,k\in\{1,2,3\}$. Based on the results $\mathrm{Tr}(\rho_{\xi_k}\hat{H}^{ij\rightarrow k})\leqslant\vert\hat{H}^{ij\rightarrow k}\vert^{\max}$, $\mathrm{Tr}(\rho_{\xi_k}\hat{H}^{k\rightarrow ij})\leqslant\vert\hat{H}^{k\rightarrow ij}\vert^{\max}$, $\int d\xi_k\Omega_k(\xi_k)=1$, and $\sum_{k=1}^6q_k=1$, an LSI for multipartite case can be obtained
\begin{equation}
\langle\hat{H}\rangle\leqslant \beta_{\mathrm{GMST}}.
\end{equation}
If this inequality is violated, we say that the state is genuine multipartite two-way steerable.
 Finally, if $\beta_{\mathrm{GMST}}$ can be obtained with the compatible  measurements performed by Alice, we say that the $\beta_{\mathrm{GMST}}$ is an LHS-attainable threshold.

\section{applications}
\label{Sec5}

In the section above, we have developed a protocol  for detecting GMST with LSIs. In this section, some explicit examples will be provided for constructing the multipartite LSI from a given operator.

\subsection{The Svetlichny  operator}
With the conventional denotation $ABC=\hat{A}\otimes \hat{B}\otimes\hat{C}$,  the Svetlichny operator is~\cite{svet}
\begin{eqnarray}
\label{Svt}
\hat{\mathcal{B}}_{\mathrm{Sve}}&=&A_1B_1C_1+A_1B_1C_2+ A_2B_1C_1- A_2B_1C_2\nonumber\\
&+& A_1B_2C_1- A_1B_2C_2- A_2B_2C_1 -A_2B_2C_2,\nonumber
\end{eqnarray}
where all the operators $A_i, B_j$, and $C_k$ are $2$-value operators. The original Svetlichny inequality, which is designed to detect the GMNL, takes the form: $\langle \hat{\mathcal{B}}_{\mathrm{Sve}}\rangle\leqslant \beta_{\mathrm{GMNL}}=4$.

First, let us consider the case that particle 1 is on Alice's side while the rest two are on Bob's side. Using Eq.~\eqref{ddddd},
the operator
$\hat{H}^{1\rightarrow 23}$ becomes
\begin{eqnarray}
\hat{H}^{1\rightarrow 23}&=&\mathfrak{d}(0\vert\xi)(B_1C_1+ B_1C_2+ B_2C_1- B_2C_2)\nonumber\\
&+&\mathfrak{d}(1\vert\xi)(B_1C_1- B_1C_2- B_2C_1- B_2C_2).\nonumber
\end{eqnarray}
According to Eq.~\eqref{hhhh}, there are four terms, $\hat{H}_{00}=2(B_1C_1-B_2C_2)$, $\hat{H}_{01}=2(B_1C_2+B_2C_1)$, $\hat{H}_{10}=-2(B_1C_2+B_2C_1)$, and $\hat{H}_{11}=2(-B_1C_1+B_2C_2)$. Note that the operator $B_jC_k$ is still a $2$-value operator, and then $\vert \hat{H}_{ab}\vert ^{\max}\leqslant4,\forall a, b\in \{0,1\}$. Therefore,  $\vert\hat{H}^{1\rightarrow 23}\vert^{\max}\leqslant4$.

Second, we consider that particles 1 and 2 are in Alice's hand while particle 3 is in Bob's hand. With $\hat{\Pi}^{0}_1-\hat{\Pi}^{1}_{1}=A_1B_1$, $\hat{\Pi}^{0}_2-\hat{\Pi}^{1}_{2}=A_1B_2$,   $\hat{\Pi}^{0}_3-\hat{\Pi}^{1}_{3}=A_2B_1$, and  $\hat{\Pi}^{0}_4-\hat{\Pi}^{1}_{4}=A_2B_2$, from  Eq.~\eqref{ddddd}, there is
\begin{eqnarray}
\hat{H}^{12\rightarrow 3}&=&\mathfrak{d}(1\vert\xi)(C_1+C_2)+\mathfrak{d}(2\vert\xi)(C_2-C_1) \nonumber\\
&+&\mathfrak{d}(3\vert\xi)(C_1-C_2)-\mathfrak{d}(4\vert\xi)(C_1+C_2),\nonumber
\end{eqnarray}
and according to Eq.~\eqref{hhhh}, there are sixteen  terms: $\hat{H}_{0000}=0$, $\hat{H}_{0001}=2(C_1+C_2)$, $\hat{H}_{0010}=2(C_2-C_1)$, $\hat{H}_{0011}=4C_2$, $\hat{H}_{0100}=2(C_1-C_2)$, $\hat{H}_{0101}=4C_1$, $\hat{H}_{0110}=0$, $\hat{H}_{0111}=2(C_1+C_2)$, $\hat{H}_{1000}=-2(C_1+C_2)$, $\hat{H}_{1001}=0$, $\hat{H}_{1010}=-4C_1$, $\hat{H}_{1011}=-2(C_1-C_2)$, $\hat{H}_{1100}=-4C_2$, $\hat{H}_{1101}=2(C_1-C_2)$, $\hat{H}_{1110}=-2(C_1+C_2)$, and $\hat{H}_{1111}=0$. With the definition $\vert\hat{H}^{12\rightarrow 3}\vert^{\max}=\max_{abcd}\{\vert\hat{H}_{abcd}\vert^{\max}\}$, we have $\vert\hat{H}^{12\rightarrow 3}\vert^{\max}=4$. Similarly, $\vert\hat{H}^{13\rightarrow 2}\vert^{\max}=\vert\hat{H}^{2\rightarrow 13}\vert^{\max}=4$, and $\vert\hat{H}^{23\rightarrow 1}\vert^{\max}=\vert\hat{H}^{3\rightarrow 12}\vert^{\max}\leqslant4$. According to Eq.~\eqref{betamax}, we can obtain $\beta_{\mathrm{GMST}}=4$. The LSI constructed from the Svetlichny operator is
\begin{equation}
\label{svtlsi}
\langle \hat{\mathcal{B}}_{\mathrm{Sve}}\rangle\leqslant \beta_{\mathrm{GMST}}=4.
\end{equation}

The maximum value, $\langle \hat{\mathcal{B}}_{\mathrm{Sve}}\rangle_{\max}=4\sqrt{2}$, can be attained for the Greenberger-Horne-Zeilinger (GHZ) state~\cite{GHZstate}, $\vert\Phi_\mathrm{{GHZ}}\rangle=(\vert 000\rangle+\vert 111\rangle)/\sqrt{2}$, with the experiment settings: $A_1=\sigma_x$, $A_2=\sigma_y$, $B_1=(\sigma_x-\sigma_y)/\sqrt{2}$, $B_2=(\sigma_x+\sigma_y)/\sqrt{2}$, $C_1=\sigma_x$, and $C_2=-\sigma_y$. If the compatible measurement in Eq.~\eqref{MUBs} is performed by Alice on particle 1: $\hat{\Pi}^a_1=(I_2+(-1)^a\sigma_x/\sqrt{2})/2$, $\hat{\Pi}^b_2=(I_2+(-1)^b\sigma_y/\sqrt{2})/2$, and the settings for particle 2 and 3 keep unchanged, one can have $\langle \hat{\mathcal{B}}_{\mathrm{Sve}}\rangle=4$. From it, we know that the steering threshold in Eq.~\eqref{svtlsi} is LHS-attainable.

\subsection{The Mermin operator}
With the usually used denotation $X=\sigma_x$, $Y=\sigma_y$, and $Z=\sigma_z$, the Mermin operator can be introduced~\cite{Mermin},
\begin{equation}
\hat{\mathcal{B}}_{\mathrm{Mer}}=X_1X_2X_3-X_1Y_2Y_3-Y_1X_2Y_3-Y_1Y_2X_3.
\end{equation}
For the case in which particle 1 is on Alice's side and particles 2 and 3 are on Bob's side, the operator $\hat{H}^{1\rightarrow 23}$ can be given
\begin{eqnarray}
\hat{H}^{1\rightarrow 23}&=&\mathfrak{d}(1\vert\xi)(X_2X_3-Y_2Y_3)\nonumber\\
&-&\mathfrak{d}(2\vert\xi)(X_2Y_3+Y_2X_3).\nonumber
\end{eqnarray}
 According to Eq.~\eqref{hhhh}, there are four terms
\begin{eqnarray}
\label{mermin}
\hat{H}_{00}&=&X_2X_3-Y_2Y_3- X_2Y_3- Y_2X_3,\nonumber\\
\hat{H}_{01}&=&X_2X_3 -Y_2Y_3 +X_2Y_3 +Y_2X_3,\nonumber\\
\hat{H}_{10}&=&-X_2X_3 +Y_2Y_3 -X_2Y_3- Y_2X_3,\\
\hat{H}_{11}&=&-X_2X_3 +Y_2Y_3 +X_2Y_3+ Y_2X_3.\nonumber
\end{eqnarray}
With a simple calculation, $\vert \hat{H}_{bc}\vert^{\max}=2\sqrt{2},\ \forall b,c \in\{1,2\}$, and therefore, $\vert\hat{H}^{1\rightarrow 23}\vert^{\max}=2\sqrt{2}$. Via the similar derivation of the Svetlichny operator, there is $\vert\hat{H}^{12\rightarrow 3}\vert^{\max}=2\sqrt{2}$. The Mermin operator is invariant under relabelling the sites of the particles, and one can obtain $\vert\hat{H}^{1\rightarrow 23}\vert^{\max}=\vert\hat{H}^{2\rightarrow 13}\vert^{\max}=\vert\hat{H}^{3\rightarrow 12}\vert^{\max}$ and $\vert\hat{H}^{23\rightarrow 1}\vert^{\max}=\vert\hat{H}^{13\rightarrow 2}\vert^{\max}=\vert\hat{H}^{12\rightarrow 3}\vert^{\max}$. Collecting the results above together, there is $\beta_{\mathrm{GMST}}=2\sqrt{2}$. Finally, the LSI, where the Mermin operator is applied for detecting GMST, can be expressed as
\begin{equation}
\label{merlsi}
\langle\hat{\mathcal{B}}_{\mathrm{Mer}}\rangle\leqslant\beta_{\mathrm{GMST}}=2\sqrt{2}.
\end{equation}

Here, it should be noted that the original Mermin inequality, $\langle\hat{\mathcal{B}}_{\mathrm{Mer}}\rangle\leqslant\beta_{\mathrm{NL}}=2$, is designed for detecting nonlocality for three-particle system, while the Svetlichny inequality is designed for detecting the genuine three-partite nonlocality. For the Mermin operator, the steering threshold ($\beta_{\mathrm{GMST}}=2\sqrt{2}$) is not equal to the nonlocal threshold ($\beta_{\mathrm{NL}}=2$).

\subsection{The GHZ-type operator}
As the third example,  we consider the operator
\begin{eqnarray}
\label{GHZ}
\hat{H}_{\mathrm{GHZ}}&=&X_1X_2X_3-X_1Y_2Y_3-Y_1X_2Y_3-Y_1Y_2X_3\nonumber\\
&+&Z_1I_2Z_3+I_1Z_2Z_3+Z_1Z_2I_3,
\end{eqnarray}
with $I_k$ the identity operator for particle $k$. The subscript GHZ indicates the maximum value of the operator is obtained from the GHZ state, say $\langle\hat{H}_{\mathrm{GHZ}}\rangle_{\max}=7.$

At first, we start with the case where particle 1 is on Alice's side.  According to Eq.~\eqref{1123}, the operator $\hat{H}^{1\rightarrow 23}$ becomes
\begin{eqnarray}
\hat{H}^{1\rightarrow 23}&=&\mathfrak{d}(1\vert\xi)(X_2X_3-Y_2Y_3)\nonumber\\
&-&\mathfrak{d}(2\vert\xi)(X_2Y_3+Y_2X_3).\nonumber\\
&+&\mathfrak{p}(0\vert 3,\xi)(Z_2Z_3+Z_2I_3+I_2Z_3)\nonumber\\
&+&\mathfrak{p}(1\vert 3,\xi)(Z_2Z_3-Z_2I_3-I_2Z_3),\nonumber
\end{eqnarray}
and according to Eq.~\eqref{hhhh}, there are eight terms,
\begin{eqnarray}
\hat{H}_{ab0}&=&\hat{H}_{ab}+Z_2Z_3+Z_2I_3+I_2Z_3,\nonumber\\
\hat{H}_{ab1}&=&\hat{H}_{ab}+Z_2Z_3-Z_2I_3-I_2Z_3,\nonumber
\end{eqnarray}
where $a,b=1,2$ and $\hat{H}_{ab}$ is given in Eq.~\eqref{mermin}. With the standard way to calculate the maximum eigenvalue of a Hermitian operators, one can obtain $\vert \hat{H}_{abc}\vert^{\max}=1+2\sqrt{3}$, $\forall a,b,c\in\{0,1\}$. Certainly, $\vert\hat{H}^{1\rightarrow 23}\vert^{\max}=1+2\sqrt{3}$.

Second, we consider the case where the particles 1 and  2 is in Alice's hand. There are seven measurements performed by Alice, and let us arrange these measurements in sequence, $ X_1X_2, X_1Y_2, Y_1X_2, Y_1Y_1, Z_1I_2, I_1Z_2$, and $Z_1Z_2$. Using Eq.~\eqref{ddddd}, one can formally have
\begin{equation}
\hat{H}^{12\rightarrow 3}= \sum_{\mu=1}^7 \mathfrak{d}(\mu\vert\xi)\hat{F}_{\mu},\nonumber
\end{equation}
where $\hat{F}_{\mu}$ can be given in sequence: $X$, $-Y$, $-Y$, $-X$, $Z$, $Z$, and $I$. From the
definition $\hat{H}_{k_1k_2...k_7}=\sum_{\mu=1}^7 (-1)^{k_{\mu}}\hat{F}_{\mu}$, it can be easily verified that $\vert \hat{H}_{k_1k_2...k_7}\vert^{\max}\leqslant1+2\sqrt{3}$. Moreover, based on the fact that the operator in Eq.~\eqref{GHZ} is invariant under relabelling the sites of the particles, and with known results for $\hat{H}^{1\rightarrow 23}$ and  $\hat{H}^{12\rightarrow 3}$, the LSI can be given
\begin{equation}
\label{hghz}
\langle \hat{H}_{\mathrm{GHZ}}\rangle\leqslant\beta_{\mathrm{GMST}}=1+2\sqrt{3}.
\end{equation}

\subsection{Unequal-weighted LSIs}
The above three examples are all equal-weighted, the absolute value of the coefficient for each 2-value operator, say, ABC, is 1. In general, one may design unequal-weighted LSIs from the equal-weighted ones. For example, one can design a simple one from Eq.~\eqref{GHZ}
\begin{eqnarray}
\hat{H}'_{\mathrm{GHZ}}&=&X_1X_2X_3-X_1Y_2Y_3-Y_1X_2Y_3-Y_1Y_2X_3\nonumber\\
&+&\alpha(Z_1I_2Z_3+I_1Z_2Z_3+Z_1Z_2I_3),
\end{eqnarray}
where $\alpha$ is a real coefficient.
With the  derivation given in Appendix~\ref{App2}, an LSI can be obtained
\begin{equation}
\label{Eq70}
\langle \hat{H}_{\mathrm{GHZ}}'\rangle\leqslant\beta_{\mathrm{GMST}}=\vert \alpha\vert +2\sqrt{\alpha^2+2}.
\end{equation}
By letting $\alpha=1$, the LSI in Eq. \eqref{hghz} is recovered. If $\alpha=0$, then Eq.~\eqref{merlsi} is also arrived at.  

With experiment values $(a)\equiv\langle \hat{M}_{\mathrm{Mer}}\rangle$, $(b)\equiv\langle Z_1I_2Z_3+I_1Z_2Z_3+Z_1Z_2I_3\rangle$ and $\beta_{\mathrm{GMST}}$, the optimal choice of $\alpha$ is defined to be the one which makes the ration $R\equiv[(a)+\alpha(b)]/\beta_{\mathrm{GMST}}$ has the maximum value. As an example, for the GHZ-type state,
 \begin{equation}
 \rho=V\vert\Phi_{\mathrm{GHZ}}\rangle\langle\Phi_{\mathrm{GHZ}}\vert+\frac{1-V}{8}I,\nonumber
 \end{equation}
 there are $(a)=4V$ and  $(b)=3V$. The numerical calculation give the optimal choice $\alpha=0.709$  and $R_{\max}=V/0.632$.  Now, if $V>0.632$, then the GHZ-type state is shown to be genuine three-partite steerable. The value $V>0.632$ is  better than the one $V>(1+2\sqrt{3})/7$ obtained from the LSI in Eq.~\eqref{hghz}.

Another unequal-weighted operator can be defined as
\begin{eqnarray}
\hat{H}&\equiv&\gamma(I_1I_2Z_3+Z_1I_2I_3+I_1Z_2I_3+Z_1Z_2Z_3)\nonumber\\
  &+&Z_1Z_2I_3+I_1Z_2Z_3+Z_1I_2Z_3\\
&+&\nonumber\delta(X_1X_2X_3-Y_1Y_2X_3-X_1Y_2Y_3-Y_1X_2Y_3)
\end{eqnarray}
where $\gamma\geqslant 0$. With the derivation in Appendix~\ref{App3},  one can come to
\begin{equation}
\label{Eq72}
\beta_{\mathrm{GMST}}=1+2\gamma+2\sqrt{(1+\gamma)^2+2\delta^2}.
\end{equation}
Let $(a)=\langle I_1I_2Z_3+Z_1I_2I_3+I_1Z_2I_3+Z_1Z_2Z_3\rangle$, $(b)=\langle Z_1Z_2I_3+I_1Z_2Z_3+Z_1I_2Z_3\rangle$, and $(c)=\langle X_1X_2X_3-Y_1Y_2X_3-X_1Y_2Y_3-Y_1X_2Y_3\rangle$, the expectation of $\hat{H}$ becomes more compact: $\langle \hat{H}\rangle=\gamma(a)+(b)+\delta (c)$. The free parameters, $\gamma$ and $\delta$, may be chosen according to the experimental data, $(a)$, $(b)$, and $(c)$.

Let us consider the generalized GHZ state
\begin{equation}
\vert\Phi\rangle=\cos\frac{\omega}{2}\vert 000\rangle+\sin\frac{\omega}{2}\vert 111\rangle,\nonumber
\end{equation}
with $0<\omega\leqslant\pi/2$, and the expectations can be calculated, $(a)=4\cos\omega$, $(b)=3$, and $(c)=4\sin\omega$. Now, one can make such a choice  that
\begin{equation}
\label{gamma}
\gamma=\cos{\omega},\delta=\sin\omega,
\end{equation}
and a fixed expectation $\langle \hat{H}\rangle=7$ can be obtained. Under the choices in Eq.~\eqref{gamma}, the threshold $\beta_{\mathrm{GMST}}$ is a function of $\omega$,
\begin{equation}
\beta_{\mathrm{GMST}}(\omega)=1+2\cos\omega+2\sqrt{(1+\cos\omega)^2+2\sin^2\omega}.\nonumber
\end{equation}
Setting $\d\beta_{\mathrm{GMST}}(\omega)/\d\omega=0$, one can obtain the maximal value of $\beta_{\mathrm{GMST}}(\omega)$, $\max_{\omega}{\beta_{\mathrm{GMST}}(\omega})=7$, when $\omega=0$. Therefore, in the parameter range $0< \omega\leqslant\pi/2$, which is allowed for the generalized GHZ state, there always exists that
\begin{equation}
\langle \hat{H}\rangle> \beta_{\mathrm{GMST}}(\omega).\nonumber
\end{equation}
Thus, we conclude that the generalized GHZ state is genuine three-partite two-way steerable.

\subsection{Arbitrary $N$-particle case}
The above examples are all about three-particle cases. For the general $N$-particle case, the derivation of LSI usually becomes very tedious as $N$ is increasing. However, if the operator
\begin{equation}
\label{localde}
\hat{H}=\sum_{k_1,k_2,...,k_n} \gamma(k_1k_2...k_n)\bigotimes_{j=1}^N\hat{A}^{j}_{k_j},
\end{equation}
keeps unchanged under relabelling the sites of the particles, the derivation of the LSI may become simplified. For such a completely symmetric $\hat{H}$, considering the case that the particles $1$, $2$, ..., and $m$ are on Alice's side, and the rest ones are on Bob's side, an operator $\hat{H}^m(\xi)$ can be introduced and the one-way steering threshold from Alice to Bob can be calculated with $\hat{H}^m(\xi)$, $\beta_{\mathrm{ST}}^{\mathrm{A}\rightarrow \mathrm{B}}=\vert\hat{H}^m(\xi)\vert^{\max}$. If all the calculations have been completed, the steering
threshold $\beta_{\mathrm{GMST}}$ is also obtained,
\begin{equation}
\label{decided}
\beta_{\mathrm{GMST}}=\max_{m\in\{1,2,...,N-1\}}\vert\hat{H}^m(\xi)\vert^{\max}.
\end{equation}

As a concrete example, let us consider the operator
\begin{equation}
\hat{H}_{\mathrm{NGHZ}}=\bigotimes_{j=1}^N\left(\vert 0\rangle_j\langle 1\vert\right)+\bigotimes _{j=1}^N\left(\vert 1\rangle_j\langle 0\vert\right),
\end{equation}
which is invariant under the relabeling the sites of particles. The subscript indicates that maximal expectation of the operator, $\langle \hat{H}_{\mathrm{NGHZ}}\rangle_{\max}=1$, can be attained for the $N$-particle GHZ state, $\vert\Phi_{\mathrm{NGHZ}}\rangle=(\otimes^N_{j=1}\vert 0\rangle_j+\otimes^N_{j=1}\vert 1\rangle_j)/\sqrt{2}$, with $\vert 0\rangle_j$ and $\vert 1\rangle_j$ the basis vectors for the $j$th particle. Certainly, using the simple relations
\begin{equation}
\label{simplera}
\vert 0\rangle\langle 1\vert=\frac{1}{2}(\sigma_x+\im\sigma_y),
\vert 1\rangle\langle 0\vert=\frac{1}{2}(\sigma_x-\im\sigma_y),
\end{equation}
$\hat{H}_{\mathrm{NGHZ}}$ can be expressed as the form in Eq.~\eqref{localde}. Now, let us introduce the following operators
\begin{eqnarray}
\label{fff}
\hat{A}^m_1&=&\bigotimes_{j=1}^m\left(\vert 0\rangle_j\langle 1\vert\right)+\bigotimes_{j=1}^m\left(\vert 1\rangle_j\langle 0\vert\right),\nonumber\\
\hat{A}^m_2&=&\im\bigotimes_{j=1}^m\left(\vert 0\rangle_j\langle 1\vert\right)-\im\bigotimes_{j=1}^m\left(\vert 1\rangle_j\langle 0\vert\right),\nonumber\\
\hat{F}^{N-m}_1&=&\frac{1}{2}\left[\bigotimes_{j=m+1}^N\left(\vert 0\rangle_j\langle 1\vert\right)+\bigotimes_{j=m+1}^N \left(\vert 1\rangle_j\langle 0\vert\right)\right],\nonumber\\
\hat{F}^{N-m}_2&=&-\frac{\im}{2}\left[\bigotimes_{j=m+1}^N\left(\vert 0\rangle_j\langle 1\vert\right)-\bigotimes_{j=m+1}^N \left(\vert 1\rangle_j\langle 0\vert\right)\right],
\end{eqnarray}
and $\hat{H}_{\mathrm{NGHZ}}$ can be decomposed as
\begin{equation}
\label{finalH}
\hat{H}_{\mathrm{NGHZ}}=\sum_{\mu=1}^2 \hat{A}_{\mu}^{m}\otimes \hat{F}^{N-m}_{\mu}.
\end{equation}
With the decomposition above, assuming that each $\hat{A}^m_{\mu}$ can be expanded with a set of 2-value operators $\{\hat{B}_{x}\}$ as $\hat{A}^m_{\mu}=\sum_x c_{\mu}^x\hat{B}_x$, and replacing each $\hat{B}_{x}$ with $\mathfrak{d}(x\vert \xi)$, we
define the two functions
\begin{equation}
\label{functionxi}
f_{\mu}(\xi)=\sum_x c_{\mu}^x\mathfrak{d}(x\vert \xi),\mu=1,2,
\end{equation}
with $\mathfrak{d}(x\vert \xi)\in\{1,-1\}$. The operator $\hat{H}^m(\xi)$ can be expressed as $\hat{H}^m(\xi)=\sum_{\mu=1}^2 f_{\mu}(\xi)\hat{F}^{N-m}_{\mu}$. The maximum eigenvalue
for $\hat{H}^m(\xi)$ can be derived
\begin{equation}
\label{eigenvalue}
\vert\hat{H}^m(\xi)\vert^{\max}=\max_{\vert\phi\rangle, \mathfrak{d}(x\vert \xi)}
\langle \phi\vert\hat{H}^m(\xi)\vert\phi\rangle.
\end{equation}
It can be noted that the operator $\hat{F}^{N-m}_1$ has only two non-zero eigenvalues, $\pm1/2$, with the corresponding eigenvectors
\begin{equation}
\vert \phi_{\pm}\rangle=\frac{\sqrt{2}}{2}\left(\bigotimes_{j=m+1}^N\vert 0\rangle_j\pm\bigotimes_{j=m+1}^N \vert 1\rangle_j\right).\nonumber
\end{equation}
Obviously, the nonzero eigenvalues of $\hat{F}^{N-m}_2$  are also $\pm1/2$ with the corresponding eigenvectors
\begin{equation}
\vert \psi_{\pm}\rangle=\frac{\sqrt{2}}{2}\left(\bigotimes_{j=m+1}^N\vert 0_j\rangle \pm \im\bigotimes_{j=m+1}^N \vert 1_j\rangle\right).\nonumber
\end{equation}
To obtain $\vert\hat{H}^m(\xi)\vert^{\max}$, it is only required to design $\vert\phi\rangle$ with a two-parameter model
\begin{equation}
\vert\phi\rangle=\cos\frac{\theta}{2}\bigotimes_{j=m+1}^N\vert 0\rangle_j+e^{-\im\phi}\sin\frac{\theta}{2}\bigotimes_{j=m+1}^N \vert 1\rangle_j.\nonumber
\end{equation}
Simple algebra shows that $\langle \phi\vert\hat{F}^{N-m}_1\vert\phi\rangle=\sin\theta\cos\phi/2$ and $\langle \phi\vert\hat{F}^{N-m}_2\vert\phi\rangle=\sin\theta\sin\phi/2$.  The optimal choice of $\theta$ is $\theta=\pi/2$. Now, the Eq.~\eqref{eigenvalue} can be simplified as
\begin{equation}
\label{simplified}
2\vert\hat{H}^m(\xi)\vert^{\max}=\max_{\phi, \mathfrak{d}(x\vert \xi)}(\cos\phi f_1(\xi)+
\sin\phi f_2(\xi))
\end{equation}
Before proceeding the derivation, an interpretation for the operators in Eq.~\eqref{fff} can be given first. Now, we define
a state
\begin{equation}
\label{psim}
\vert\Psi^m\rangle=\frac{1}{\sqrt{2}}\left(\bigotimes_{j=1}^m\vert 0\rangle_j+\bigotimes_{j=1}^m\vert 1\rangle_j\right),
\end{equation}
which is nothing else but the GHZ state for $m\geqslant 3$.  The operator $\vert \Psi^m\rangle\langle \Psi^m\vert$ can be decomposed into two terms: The diagonal term $\left(\bigotimes_{j=1}^m \vert 0\rangle_j\langle 0\vert+\bigotimes_{j=1}^m\vert 1\rangle_j\langle 1\vert\right)/2$, and the off-diagonal term $\left(\bigotimes_{j=1}^m \vert 0\rangle_j\langle 1\vert+\bigotimes_{j=1}^m \vert 1\rangle_j\langle 0\vert\right)/2$. The off-diagonal term can also be expressed as $\hat{A}^m_1/2$. Meanwhile, another  state can be introduced as
\begin{equation}
\label{phim}
\vert\Phi^m\rangle=\frac{1}{\sqrt{2}}\left(\bigotimes_{j=1}^m\vert 0\rangle_j+\im\bigotimes_{j=1}^m\vert 1\rangle_j\right),
\end{equation}
which is also a GHZ state for $m\geqslant 3$. The off-diagonal term for $\vert \Phi^m\rangle\langle \Phi^m\vert$ can be expressed as $-\hat{A}^m_2/2$.
In 1990, Mermin introduced an operator
\begin{equation}
\hat{\mathrm{\mathcal{B}}}_{\mathrm{Mer}}=\frac{1}{2\im}\left[\bigotimes_{j=1}^m
\left(\sigma_x^j+\im\sigma_y^j\right)-\bigotimes_{j=1}^m\left(\sigma_x^j-\im\sigma_y^j\right)\right],
\end{equation}
with $\sigma_x^j$ and $\sigma_y^j$ the Pauli matrices for the $j$th particle, and it was showed that the Mermin's inequality is maximally violated by the states in Eq.~\eqref{phim} (for $m\geqslant 3$)~\cite{Mermin}. When $m\geqslant 3$, with the simple relations in Eq.~\eqref{simplera}, one can have $\hat{\mathrm{\mathcal{B}}}_{\mathrm{Mer}}=-2^{m-1}\hat{A}^m_2$. Usually, the Mermin's operator can also be defined as $\hat{\mathrm{\mathcal{B}}}_{\mathrm{Mer}}=2^{m-1}\hat{A}^m_1$, and the Mermin's inequality is maximally violated by the state in Eq.~\eqref{psim}. In Eq. \eqref{fff}, the definitions of $\hat{F}^m_{\mu}$ are very similar to the ones for $\hat{A}^m_{\mu}$.

According to the operators in Eq.~\eqref{fff}, one can construct $\hat{A}^m_{\mu}$ from $\hat{A}_{\mu}^{m-1}$ via a following way:
\begin{eqnarray}
\hat{A}^m_{1}&=&\frac{1}{2}(\hat{A}_1^{m-1}\otimes\sigma_x+\hat{A}_2^{m-1}\otimes\sigma_y),
\nonumber\\
\label{construct}
\hat{A}^m_{2}&=&\frac{1}{2}(-\hat{A}_1^{m-1}\otimes\sigma_y+\hat{A}_2^{m-1}\otimes\sigma_x).
\end{eqnarray}
Let us start from $A^1_1=\sigma_x$ and $\hat{A}^1_2=-\sigma_y$, and there are $\hat{A}^2_1=(\sigma_x\otimes\sigma_x-\sigma_y\otimes \sigma_y)/2$ and $\hat{A}^2_2=(\sigma_x\otimes\sigma_y+\sigma_y\otimes \sigma_x)/2$. Using Eq. \eqref{construct} again, one can obtain $\hat{A}^3_1=(\sigma_x\otimes\sigma_x\otimes\sigma_x-\sigma_y\otimes\sigma_y\otimes\sigma_x-\sigma_x\otimes\sigma_y\otimes\sigma_y-\sigma_y\otimes\sigma_x\otimes\sigma_y)/4$ and $\hat{A}^3_2=-(\sigma_x\otimes\sigma_x\otimes\sigma_y-\sigma_y\otimes\sigma_y\otimes\sigma_y+\sigma_x\otimes\sigma_y\otimes\sigma_x+\sigma_y\otimes\sigma_x\otimes\sigma_x)/4$. Now, let $S^{m=3}_1\equiv\{\sigma_x\otimes\sigma_x\otimes\sigma_x, \sigma_y\otimes\sigma_y\otimes\sigma_x,\sigma_x\otimes\sigma_y\otimes\sigma_y,\sigma_y\otimes\sigma_x\otimes\sigma_y\}$ be the complete set of the 2-value operators appeared in $\hat{A}^3_1$ with non-zero coefficients, and $S^{m=3}_2\equiv\{\sigma_x\otimes\sigma_x\otimes\sigma_y, \sigma_y\otimes\sigma_y\otimes\sigma_y,\sigma_x\otimes\sigma_y\otimes\sigma_x,\sigma_y\otimes\sigma_x\otimes\sigma_x\}$ is the complete set for the ones in $\hat{A}^3_2$. The definitions for $S^{m=3}_{\mu}$ $(\mu=1,2)$  can be easily generalized to the cases with arbitrary $m$. For two arbitrary 2-value operators $\hat{G}\in S^{m=3}_1$ and $\hat{K}\in S^{m=3}_2$, with the explicit definitions for $S^{m=3}_{\mu}$ above, one can easily verify that $\hat{G}\neq \hat{K}$. With Eq. \eqref{construct}, one can have that the operators in $S^{m=4}_1$ take the form $\pm\hat{G}\otimes\sigma_x$ or $\pm\hat{K}\otimes \sigma_y$, while the operators in $S^{m=4}_2$ are $\pm\hat{K}\otimes \sigma_x$ or $\pm\hat{G}\otimes \sigma_y$. Since $\hat{G}\neq \hat{K}$, one can conclude that the two sets, $S^{m=4}_1$ and $S^{m=4}_2$, do not share a common 2-value operator. For the same reason, one can conclude that the operators in $S^m_1 (S^m_2)$ are independent from the ones in $S^m_2(S^m_1)$ as $m$ increases.

Formally, a single index $x$ is used to label the 2-value operators in the set $S^m_1$, and $S^m_1\equiv\{\hat{G}_x\}$. From Eq.~\eqref{construct}, $\hat{A}^m_1$ can be rewritten as $\hat{A}^m_1=\sum_{x=1}^{2^{m-1}}c_x\hat{G}_x/2^{m-1}$ with $\vert c_x\vert=1$. Meanwhile, the index $y$ is used to label the 2-value operators in the set $S^m_2$, $S^m_2\equiv\{\hat{K}_y\}$, and $\hat{A}^m_2$ can be rewritten as $\hat{A}^m_2=\sum_{y=1}^{2^{m-1}}\bar{c}_y\hat{K}_y/2^{m-1}$ with $\vert \bar{c}_y\vert=1$.  From Eq.~\eqref{functionxi}, two functions can be obtained as follows
\begin{eqnarray}
f_1(\xi)&=&\frac{1}{2^{m-1}}\sum_{x=1}^{2^{m-1}}c_x \mathfrak{d}(x\vert \xi),\ \mathrm{with}\ \vert c_x\vert=1,  \nonumber\\
\label{newfifi}
f_2(\xi)&=&\frac{1}{2^{m-1}}\sum_{y=1}^{2^{m-1}}\bar{c}_y\mathfrak{d}(y\vert \xi),\ \mathrm{with}\ \vert\bar{c}_y\vert=1,
\end{eqnarray}
with $\mathfrak{d}(x\vert \xi)\in\{1,-1\}$ and $\mathfrak{d}(y\vert \xi)\in\{1,-1\}$.

Since the two sets $S^m_1$ and $S^m_2$ do not share any common  2-value operator, the two sets of parameters $\{\mathfrak{d}(x\vert \xi)\}$ (in $f_1(\xi)$) and $\{\mathfrak{d}(y\vert \xi)\}$ (in $f_2(\xi)$) should be independent from each other, and Eq.~\eqref{simplified} becomes
\begin{equation}
\label{f1f2}
2\vert\hat{H}^m(\xi)\vert^{\max}=\max_{\phi}\left[\cos\phi\max_{\mathfrak{d}(x\vert \xi)}f_1(\xi)+\sin\phi\max_{\mathfrak{d}(y\vert \xi)}f_2(\xi)\right].\nonumber
\end{equation}
With the expressions in Eq.~\eqref{newfifi}, one can have $\max_{\mathfrak{d}(x\vert \xi)}f_1(\xi)=\max_{\mathfrak{d}(y\vert \xi)}f_2(\xi)=1$ and obtain the optimal choice of $\phi$ satisfying $\cos\phi=\sin\phi=1/\sqrt{2}$. In Eq.~\eqref{newfifi}, although each $f_{\mu}(\xi)$ has a number of $2^{m-1}$ free parameters $\mathfrak{d}(x\vert \xi)$ (or $\mathfrak{d}(y\vert \xi)$, it also contains a factor $1/2^{m-1}$, and this is the reason why $\max_{\mathfrak{d}(x\vert \xi)}f_1(\xi)$ and $\max_{\mathfrak{d}(y\vert \xi)}f_2(\xi)$ keep unchanged when $m$ is increasing. Finally, one can obtain a relation, $\vert\hat{H}^m(\xi)\vert^{\max}=1/\sqrt{2}$, which is independent of the actual value of $m$. According to the definition in Eq.~\eqref{decided}, an LSI for detecting the genuine $N$-partite two-way steerability is arrived at
\begin{equation}
\label{arbitraryN}
\langle\hat{H}_{\mathrm{NGHZ}}\rangle\leqslant\beta_{\mathrm{GMST}}=\frac{\sqrt{2}}{2},
\end{equation}
where it is required that $\hat{H}_{\mathrm{NGHZ}}$ should be expressed as the standard form in Eq.~\eqref{localde}. If $N=2$, the inequality becomes $\langle \sigma_x\otimes \sigma_x\rangle-\langle \sigma_y\otimes\sigma_y\rangle \leqslant \sqrt{2}$. Obviously, it  is similar to the LSI in Eq.~\eqref{lsi1}. For the case $N=3$, it can be easily verified that the criterion in Eq.~\eqref{arbitraryN} is equivalent to the two-way LSI in Eq.~\eqref{merlsi} constructed from Mermin's operator, $\hat{\mathcal{B}}_{\mathrm{mer}}=4\hat{A}^3_1$.

It should be emphasized that the ways to deal with the multipartite steering are not unique~\cite{rmd}. The way in this work belongs to local steering introduced by He and Reid~\cite{he-reid}.  As shown in Eq.~\eqref{hermitian}, for the three parties case with a given partition ($AB\vert C$) and the untrusted parties (A and B), only local measurements are allowed. Besides the local steering, there are other two types of multipartite steering, the global steering and reduced steering~\cite{rmd}.

For the three spin-$1/2$ particle system, 2-value operators are widely used. However, if the dimension of local system is greater than two, one shall encounter the general Hermitian operator which is not a 2-value operator. In the end of this section, a way to deal with the general operators is suggested with a simple example where the global steering is considered.

In the global  steering, the untrusted parties can perform global measurement. Consider the global steering from Alice to Bob, and for the decomposition of $\hat{H}_{\mathrm{NGHZ}}$ in Eq.~\eqref{finalH}, each $\hat{A}^m_{\mu}$ is no-longer a 2-value operator if $m\geqslant2$. In the $d$-dimensional system with $d=2^m$, the eigenvalues for each $\hat{A}^m_{\mu}$ are $\pm 1$ and 0 (with a number of $d-2$ zeros). Denote the eigenvectors corresponding two nonzero eigenvalue of $\hat{A}^m_1$ by $\vert \phi^1_1\rangle$ and $\vert \phi^2_1\rangle$, say, $\hat{A}^m_1\vert \phi^1_1\rangle=\vert \phi^1_1\rangle$ and $\hat{A}^m_1\vert \phi^2_1\rangle=-\vert \phi^2_1\rangle$, and one can define
\begin{equation}
\label{Pi0}
\hat{\Pi}^{k}_1=\vert \phi^k_1\rangle\langle\phi^k_1\vert\ (k=1,2),\ \ \hat{\Pi}^0_1=I_d-\sum_{k=1}^{2}\hat{\Pi}^{k}_1,
\end{equation}
and express $\hat{A}^m_1$  as $\hat{A}^m_1=\hat{\Pi}^{1}_1-\hat{\Pi}^{2}_1$. Similarly, by introducing the  eigenvectors, $\hat{A}^m_2\vert \phi^1_2\rangle=\vert \phi^1_2\rangle$ and $\hat{A}^m_2\vert \phi^2_2\rangle=-\vert \phi^2_2\rangle$, one can also define
\begin{equation}
\hat{\Pi}^{k}_2=\vert \phi^k_2\rangle\langle\phi^k_2\vert\ (k=1,2),\ \ \hat{\Pi}^0_2=I_d-\sum_{k=1}^{2}\hat{\Pi}^{k}_2,\nonumber
\end{equation}
and have the relation $\hat{A}^m_2=\hat{\Pi}^{1}_2-\hat{\Pi}^{2}_2$. Furthermore, the operator in Eq.~\eqref{finalH} can be rewritten as
\begin{equation}
\label{global}
\hat{H}_{\mathrm{NGHZ}}=\sum_{\mu=1}^2(\hat{\Pi}^1_{\mu}-\hat{\Pi}^2_{\mu})
\otimes \hat{F}^{N-m}_{\mu}.
\end{equation}
The operators $\hat{\Pi}^a_{\mu}$ above can be viewed as the global measurements, and obviously, these operators  do not satisfy the local-measurement requirement in Eq.~\eqref{hermitian}. For the steering from Alice to Bob, each $\hat{\Pi}^k_{\mu}$ with $k\in\{0,1,2\}$ should be replaced with its corresponding $\mathfrak{p}(k\vert\mu,\xi)$ under the condition $\sum_{k=0}^{2}\mathfrak{p}(k\vert\mu,\xi)=1$. In the deterministic model, $\mathfrak{p}(k\vert\mu,\xi)\in\{0,1\}$, there are altogether three situations: (a) $\mathfrak{p}(0\vert \mu,\xi)=1, \mathfrak{p}(1\vert \mu,\xi)=\mathfrak{p}(2\vert \mu,\xi)=0$; ({b}) $\mathfrak{p}(1\vert \mu,\xi)=1,\mathfrak{p}(0 \vert \mu,\xi)=\mathfrak{p}(2\vert \mu,\xi)=0$; and (c) $\mathfrak{p}(2\vert \mu,\xi)=1, \mathfrak{p}(0 \vert \mu,\xi)=\mathfrak{p}(1\vert \mu,\xi)=0$. Now, for the global steering from Alice to Bob, there are nine possible $\hat{H}(\xi)$, and besides the trivial one $\hat{H}_{00}=0$, these operators are listed as follows
\begin{eqnarray}
\hat{H}_{01}&=&\hat{F}^{N-m}_2,\ \ \ \ \ \ \ \ \ \ \ \ \ \ \ \ \ \ \ \ \ \hat{H}_{02}= -\hat{F}^{N-m}_2,\nonumber\\
\hat{H}_{10}&=&\hat{F}^{N-m}_1,\ \ \ \ \ \ \ \ \ \ \ \ \ \ \ \ \ \ \ \ \ \hat{H}_{20}= -\hat{F}^{N-m}_1,\nonumber\\
\hat{H}_{11}&=&\hat{F}^{N-m}_1+\hat{F}^{N-m}_2,\ \ \ \ \hat{H}_{12}=\hat{F}^{N-m}_1-\hat{F}^{N-m}_2,\nonumber\\
\hat{H}_{21}&=&-\hat{F}^{N-m}_1+\hat{F}^{N-m}_2,\ \hat{H}_{22}=-\hat{F}^{N-m}_1-\hat{F}^{N-m}_2.\nonumber
\end{eqnarray}
By some algebra, $\vert\hat{H}_{01}\vert^{\max}=\vert\hat{H}_{02}\vert^{\max}=\vert\hat{H}_{10}\vert^{\max}=\vert\hat{H}_{20}\vert^{\max}=1/2$, and  $\vert\hat{H}_{11}\vert^{\max}=\vert\hat{H}_{12}\vert^{\max}=\vert\hat{H}_{21}\vert^{\max}=\vert\hat{H}_{22}\vert^{\max}=\sqrt{2}/2$. From the definition, $\beta_{\mathrm{ST}}^{\mathrm{A}\rightarrow \mathrm{B}}=\max_{a,b\in\{0,1,2\}}\{ \vert\hat{H}_{ab}\vert^{\max}\}$, one can obtain the one-way steering threshold $\beta_{\mathrm{ST}}^{\mathrm{A}\rightarrow \mathrm{B}}=\sqrt{2}/2$ for the case when there are $m$ particles on Alice side. Via the similar derivation, $\beta_{\mathrm{ST}}^{\mathrm{A}\leftarrow \mathrm{B}}=\sqrt{2}/2$. Finally, one can come to a two-way criterion for the global steering, $\langle\hat{H}_{\mathrm{NGHZ}}\rangle \leqslant 1/\sqrt{2}$, where $\hat{H}_{\mathrm{NGHZ}}$ takes the form in Eq.~\eqref{global}.

\section{Conclusions}
\label{Sec6}
For a bipartite system, the steering from Alice to Bob is defined with the assemblage $\{\tilde{\rho}^a_{\mu}\}$ in Eq.~\eqref{df}.  By performing  the state tomography, Bob can decide the  assemblage. Under such a condition where  the full information about the assemblage is already known,  the so-called linear steering inequality (LSI) from the semidefinite program (SDP) can be constructed~\cite{Can1}: For a given assemblage $\{\tilde{\rho}^a_{\mu}\}$, the SDP is designed for testing whether it admits an LHS model. If $\{\tilde{\rho}^a_{\mu}\}$ demonstrates steering,  the solution of another program, which is based on the duality theory of SDPs, returns Hermitian operators $\{\hat{F}^a_{\mu}\}$ which can be used to define a steering inequality $\sum_{\mu}\sum_{a}\tr\left(\tilde{\rho}^a_{\mu}\hat{F}^a_{\mu}\right)\geqslant\beta_{\mathrm{LHS}}$ satisfied by all LHS assemblages and violated, \emph{in particular, by  $\{\tilde{\rho}^a_{\mu}\}$}. Obviously,  the LSI is a powerful tool   in detecting
the steerability.

 In this work, the LSIs  developed above can work without the full information about the assemblage.
According to the fundamental idea that a one-way steering inequality can be constructed by just considering the measurements performed by Bob~\cite{can22,sau,Joness}, a general scheme  has been developed to design  two kinds of LSIs, which can be applied either for detecting the two-way steerability of a bipartite system or for verifying the genuine multipartite two-way steerability of a multipartite system.

For the bipartite system, the two-way criterion is constructed from a pair of one-way
LSIs. Being compared with known protocols for designing one-way LSIs ~\cite{can22,sau,Joness,
Li,Zeng}, the present method is general in following two aspects: At fist, it has
been shown that the one-way LSI can be constructed with an operator containing free parameters; In second, the known LSIs in ~\cite{can22,sau,Joness,
Li,Zeng} can be formally expressed as $\langle \hat{H}\rangle \leqslant \beta^{\mathrm{A}\rightarrow \mathrm{B}}_{\mathrm{ST}}$. In this work, it has been demonstrated that the same operator $\hat{H}$
can be also applied to design another one-way LSI, $\gamma^{\mathrm{A}\rightarrow \mathrm{B}}_{\mathrm{ST}}\leqslant \langle \hat{H}\rangle$. We have given an explicit criterion, which belongs to the type of $\gamma^{\mathrm{A}\rightarrow \mathrm{B}}_{\mathrm{ST}}\leqslant \langle \hat{H}\rangle$, to detect the steerability of Werner state.

In previous works,  how to define and detect the genuine multipartite steerability (GMS),  is still an unsolved problem ~\cite{he-reid,li-chen,csan,rmm,gmdrm}. In present work, based on the two assumptions that GMS is  stronger than the genuine multipartite entanglement and GMS is  weaker than the genuine multipartite nonlocality, we have given a definition of the two-way GMS with a generalized LHS model. Several LSIs have been developed for detecting the two-way GMS.

For the multipartite system, the LSIs in this work are limited for the case where the measurement has a finite number of experimental settings. To show whether a state is genuine multipartite steerable, as it has been required in the bipartite system, the continuous experimental settings should be considered~\cite{Wiseman1}. We leave such kind of LSIs as our future works.

\acknowledgements
This work was supported by the National Natural Science Foundation of China (Grant No.~12147208), and the Fundamental Research Funds for the Central Universities (Grant No. 2682021ZTPY050).

\appendix
\section{Derivation of Eq.~\eqref{Eq31}}
\label{App1}
For the two sets of projective measurements, $\{\vert\phi^a_1\rangle\}$ and $\{\vert\phi^b_2\rangle\}$, they are related by a unitary transformation $U$ with $U_{ab}$ as its matrix elements, and
\begin{equation}
\vert \phi^a_1\rangle=U_{ab}\vert\phi^b_2\rangle+\sum_{c\neq b}U_{ac}\vert\phi^c_2\rangle.\nonumber
\end{equation}
For convenience, one may introduce two parameters, $\theta$ and $\eta$, and rewrite $U_{ab}$ as
\begin{equation}
U_{ab}=\cos\frac{\theta}{2}\exp\{i\eta\}, \cos\frac{\theta}{2}=\vert U_{ab}\vert.\nonumber
\end{equation}
With the unnormalized state $\vert\psi\rangle=\sum_{c\neq b}U_{ac}\vert\phi^c_2\rangle$, one can define a pair of orthogonal states
\begin{equation}
\vert e_1\rangle=\exp\{-i\eta\}\vert\phi^a_2\rangle, \vert e_2\rangle=\frac{\vert\psi\rangle}{\langle\psi\vert\psi\rangle}.\nonumber
\end{equation}
Certainly, $\langle e_i\vert e_j\rangle=\delta_{ij}$ with $i,j\in\{1,2\}$. Now, the two states, $\vert\phi^a_1\rangle$ and $\vert\phi^b_2\rangle$, can be expressed as
\begin{equation}
\vert\phi^{b}_2\rangle=\exp\{i\eta\}\vert e_1\rangle,\vert\phi^a_1\rangle=\cos\frac{\theta}{2} \vert e_1\rangle+\sin \frac{\theta}{2}
\vert e_2\rangle. \nonumber
\end{equation}
With the Pauli matrices, $\sigma_x=\vert e_1\rangle\langle e_2\vert+\vert e_2\rangle\langle e_1\vert$, $\sigma_y=-i\vert e_1\rangle\langle e_2\vert+i\vert e_2\rangle\langle e_1\vert$, and $\sigma_z=\vert e_1\rangle\langle e_1\vert-\vert e_2\rangle\langle e_2\vert$, one can have
\begin{equation}
\vert\phi^a_1\rangle\langle\phi^a_1\vert=
\frac{1}{2}(I_2+\cos\theta\sigma_z+\sin\theta\sigma_x),
\vert\phi^b_2\rangle\langle\phi^b_2\vert=\frac{1}{2}(I_2+\sigma_z).\nonumber
\end{equation}

For the operator in Eq. \eqref{gmubs}, when the steering from Alice to Bob is considered, one can obtain
\begin{eqnarray}
\hat{H}(\xi)&=& (1+\cos\omega)\left(\sum_{\bar{a}} \mathfrak{p}(\bar{a}\vert 1,\xi) \vert\phi^{\bar{a}}_{1}\rangle\langle\phi^{\bar{a}}_{1}\vert\right)\nonumber\\
&+&(1-\cos\omega)\left(\sum_{\bar{b}} \mathfrak{p}(\bar{b}\vert 2,\xi) \vert\phi^{\bar{b}}_{2}\rangle\langle\phi^{\bar{b}}_{2}\vert\right).\nonumber
\end{eqnarray}
Within the deterministic model, if $\mathfrak{p}({a}\vert 1,\xi)=\mathfrak{p}({b}\vert 2,\xi)=1$, the operator $\hat{H}_{ab}$ can take the form
\begin{equation}
\hat{H}_{ab}=2(\cos^2\frac{\omega}{2}\vert\phi^a_1\rangle\langle\phi^a_1\vert
 +\sin^2\frac{\omega}{2}\vert\phi^b_2\rangle\langle\phi^b_2\vert).\nonumber
\end{equation}
With the expressions of $\vert\phi^a_1\rangle\langle\phi^a_1\vert$ and $\vert\phi^b_2\rangle\langle\phi^b_2\vert$, $\hat{H}_{ab}$ can also be expressed as
\begin{equation}
\hat{H}_{ab}=I_2+(\sin^2\frac{\omega}{2}+\cos^2\frac{\omega}{2}\cos\theta)\sigma_z+
\cos^2\frac{\omega}{2}\sin\theta\sigma_x.\nonumber
\end{equation}
The quantity $\vert \hat{H}_{ab}\vert^{\max}$, which is the largest eigenvalue of $\hat{H}_{ab}$, can be derived,
\begin{eqnarray}
\vert \hat{H}_{ab}\vert^{\max}&=&1+\sqrt{ (\sin^2\frac{\omega}{2}+\cos^2\frac{\omega}{2}\cos\theta)^2+ (\cos^2\frac{\omega}{2}\sin\theta)^2}\nonumber\\
&=&1+\sqrt{\cos^2\omega+\sin^2\omega\cos^2\frac{\theta}{2}}.\nonumber
\end{eqnarray}
By jointing it with the relation $\cos\frac{\theta}{2}=\vert U_{ab}\vert$, there is
\begin{equation}
\vert \hat{H}_{ab}\vert^{\max}=1+\sqrt{\cos^2\omega+\sin^2\omega\vert U_{ab}\vert^2}.\nonumber
\end{equation}
Finally, according to the definition $\beta^{\mathrm{A}\rightarrow \mathrm{B}}=\max_{a,b}\{\vert \hat{H}_{ab}\vert^{\max}\}$, the steering threshold is
\begin{equation}
\beta^{\mathrm{A}\rightarrow \mathrm{B}}=1+\sqrt{\cos^2\omega+\sin^2\omega\vert U^{\mathrm{opt}}_{ab}\vert^2}.\nonumber
\end{equation}
where $\vert U^{\mathrm{opt}}_{ab}\vert$ has the largest value among all the possible $\vert U_{ab}\vert, \forall a,b \in\{0,1,...,d-1\}$.

\section{Derivation of Eq.~\eqref{Eq70}}
\label{App2}
 For the operator
\begin{eqnarray}
\hat{H}&=&X_1X_2X_3-X_1Y_2Y_3-Y_1X_2Y_3-Y_1Y_2X_3\nonumber\\
&+&\alpha(Z_1I_2Z_3+I_1Z_2Z_3+Z_1Z_2I_3),\nonumber
\end{eqnarray}
we consider the case where particle 1 is in Alice's side while the rest two particles are in Bob's side. With the replacement: $X_1\rightarrow \mathfrak{d}(1\vert \xi)$, $Y_1\rightarrow \mathfrak{d}(2\vert \xi)$,  $Z_1\rightarrow \mathfrak{d}(3\vert \xi)$, and $I_1\rightarrow 1$,  one can obtain
\begin{eqnarray}
\hat{H}^{1\rightarrow 23}&=&\mathfrak{d}(1\vert \xi)(X_2X_3-Y_2Y_3)-\mathfrak{d}(2\vert \xi)(X_2Y_3+Y_2X_3)\nonumber\\
&+&\alpha[\mathfrak{d}(3\vert \xi)(I_2Z_3+Z_2I_3)+Z_2Z_3].\nonumber
\end{eqnarray}
To calculate the eigenvalues, $\hat{H}^{1\rightarrow 23}$ can be rewritten as
\begin{equation}
\left(\begin{array}{cccc}
\alpha(1+2\mathfrak{d}(3\vert \xi)) & 0 & 0 & 2(\mathfrak{d}(1\vert \xi)+i\mathfrak{d}(2\vert \xi)) \\
0 & -\alpha & 0 & 0 \\
0 & 0 & -\alpha & 0 \\
2(\mathfrak{d}(1\vert \xi)-i\mathfrak{d}(2\vert \xi)) & 0 & 0 & \alpha(1-2\mathfrak{d}(3\vert \xi)) \\
\end{array}
\right),\nonumber
\end{equation}
which has four eigenvalues, $\lambda_{\pm}, \lambda_ 3$, and $\lambda_4$,
\begin{eqnarray}
\lambda_{\pm}&=&\alpha\pm2\sqrt{\sum_{j=1}^2 (\mathfrak{d}(j\vert\xi))^2+(\alpha\mathfrak{d}(3\vert\xi))^2},\nonumber\\
\lambda_3&=&\lambda_4=-\alpha.\nonumber
\end{eqnarray}
Based on the constraint that $-1\leqslant \mathfrak{d}(j\vert\xi)\leqslant 1, j\in\{1,2,3\}$, it can be verified that
\begin{equation}
\vert\hat{H}^{1\rightarrow 23}\vert^{\max}\leqslant \vert\alpha\vert+2\sqrt{2+\alpha^2}.\nonumber
\end{equation}
Considering the case where particles 1 and 2 are on Alice's hand while particle 3 is on Bob's hand, with the replacement: $X_1X_2\rightarrow \mathfrak{d}(1\vert \xi)$,
$X_1Y_2\rightarrow \mathfrak{d}(2\vert \xi)$,
$Y_1X_2\rightarrow \mathfrak{d}(3\vert \xi)$,
$Y_1Y_2\rightarrow \mathfrak{d}(4\vert \xi)$,
$Z_1I_2\rightarrow \mathfrak{d}(5\vert \xi)$,
$I_1Z_2\rightarrow \mathfrak{d}(6\vert \xi)$,
$Z_1Z_2\rightarrow \mathfrak{d}(7\vert \xi)$,
and $I_1I_2\rightarrow 1$, we shall get the operator $\hat{H}^{12\rightarrow3}$,
\begin{eqnarray}
\hat{H}^{12\rightarrow3}&=&(\mathfrak{d}(1\vert\xi)-\mathfrak{d}(4\vert\xi))X_3- (\mathfrak{d}(2\vert\xi)+\mathfrak{d}(3\vert\xi))Y_3 \nonumber \\
&+&\alpha(\mathfrak{d(}5\vert\xi)-\mathfrak{d}(6\vert\xi))Z_3+\alpha \mathfrak{d}(7\vert\xi)I_3.\nonumber
\end{eqnarray}
It has two eigenvalues,
\begin{eqnarray}
\lambda_{\pm}&=& \alpha \mathfrak{d}(7\vert\xi)\pm \{(\mathfrak{d}(1\vert\xi)-\mathfrak{d}(4\vert\xi))^2\nonumber\\
&+&(\mathfrak{d}(2\vert\xi)+\mathfrak{d}(3\vert\xi))^2+
\alpha^2(\mathfrak{d}(5\vert\xi)-\mathfrak{d}(6\vert\xi))^2\}^{\frac{1}{2}}.
\nonumber
\end{eqnarray}
With the constraint that $-1\leqslant \mathfrak{d}(j\vert\xi)\leqslant 1, j\in\{1,...,7\}$, it can be verified that
\begin{equation}
\vert\hat{H}^{12\rightarrow 3}\vert^{\max}\leqslant \vert\alpha\vert+2\sqrt{2+\alpha^2}.\nonumber
\end{equation}
Considering the fact that $\hat{H}$ is invariant under the relabelling the sites of the particles, one can have $\vert\hat{H}^{13\rightarrow 2}\vert^{\max}=\vert\hat{H}^{23\rightarrow 1}\vert^{\max}=\vert\hat{H}^{12\rightarrow 3}\vert^{\max}$ and $\vert\hat{H}^{2\rightarrow 13}\vert^{\max}=\vert\hat{H}^{3\rightarrow 12}\vert^{\max}=\vert\hat{H}^{1\rightarrow 23}\vert^{\max}$. According to the definition of $\beta_{\mathrm{GMST}}$ in
Eq. \eqref{betamax}, there is
\begin{equation}
\beta_{\mathrm{GMST}}=\vert\alpha\vert+2\sqrt{2+\alpha^2}.\nonumber
\end{equation}

\section{Derivation of Eq.~\eqref{Eq72}}
\label{App3}
In the three particles system, we define the operator
\begin{eqnarray}
\hat{H}&\equiv&\gamma(I_1I_2Z_3+Z_1I_2I_3+I_1Z_2I_3+Z_1Z_2Z_3)\nonumber\\
  &+&Z_1Z_2I_3+I_1Z_2Z_3+Z_1I_2Z_3\nonumber\\
&+&\nonumber\delta(X_1X_2X_3-Y_1Y_2X_3-X_1Y_2Y_3-Y_1X_2Y_3).\nonumber
\end{eqnarray}
At first, we consider the case where particle 1 is on Alice's side while particle 2 and particle 3 are on Bob's side. With the replacement: $X_1\rightarrow \mathfrak{d}(1\vert \xi)$, $Y_1\rightarrow \mathfrak{d}(2\vert \xi)$,  $Z_1\rightarrow \mathfrak{d}(3\vert \xi)$, and $I_1\rightarrow 1$, the operator $\hat{H}^{1\rightarrow 23}$ can be obtained
\begin{eqnarray}
\hat{H}^{1\rightarrow 23}&=&Z_2Z_3+\gamma \mathfrak{d}(3\vert\xi)( I_2I_3+ Z_2Z_3)\nonumber\\
&+&(\gamma+\mathfrak{d}(3\vert\xi))(I_2Z_3+Z_2I_3)\nonumber\\
&+&\mathfrak{d}(1\vert\xi)(X_2X_3-Y_2Y_3)\nonumber\\
&-&\mathfrak{d}(2\vert\xi)(Y_2X_3+X_2Y_3).\nonumber
\end{eqnarray}
To derive the eigenvalues, $\frac{1}{2}\hat{H}^{1\rightarrow 23}$ can be expressed in an equivalent form
\begin{equation}
\left(
  \begin{array}{cccc}
    \frac{1}{2}+\gamma+(1+\gamma)\mathfrak{d}(3\vert\xi) & 0 & 0 & \delta(\mathfrak{d}(1\vert \xi)+i\mathfrak{d}(2\vert \xi)) \\
    0 & -\frac{1}{2} & 0 & 0 \\
    0 & 0 & -\frac{1}{2} & 0 \\
    \delta(\mathfrak{d}(1\vert \xi)-i\mathfrak{d}(2\vert \xi)) & 0 & 0 & \frac{1}{2}-\gamma+(\gamma-1)\mathfrak{d}(3\vert\xi) \\
  \end{array}
\right).\nonumber
\end{equation}
Obviously, the operator $\hat{H}^{1\rightarrow 23}$ has four eigenvalues, $\lambda_1=\lambda_2=-1$,
\begin{eqnarray}
\lambda_{\pm}&=& 1+2\gamma \mathfrak{d}(3\vert\xi)\pm2\{[\gamma+\mathfrak{d}(3\vert\xi)]^2\nonumber\\
&+&\delta^2[\mathfrak{d}(1\vert\xi)^2+\mathfrak{d}(2\vert\xi)^2]\}^{\frac{1}{2}}.\nonumber
\end{eqnarray}
With the constraint,  $-1\leqslant \mathfrak{d}(j\vert\xi)\leqslant 1, j\in\{1,2,3\}$, one can obtain
\begin{equation}
\vert\hat{H}^{1\rightarrow 23}\vert^{\max}\leqslant 1+2\gamma+2\sqrt{(1+\gamma)^2+2\delta^2}.\nonumber
\end{equation}
 Considering the case where particles 1 and 2 are on Alice's hand while particle 3 is on Bob's hand, and with the replacement: $X_1X_2\rightarrow \mathfrak{d}(1\vert \xi)$,
$X_1Y_2\rightarrow \mathfrak{d}(2\vert \xi)$,
$Y_1X_2\rightarrow \mathfrak{d}(3\vert \xi)$,
$Y_1Y_2\rightarrow \mathfrak{d}(4\vert \xi)$,
$Z_1I_2\rightarrow \mathfrak{d}(5\vert \xi)$,
$I_1Z_2\rightarrow \mathfrak{d}(6\vert \xi)$,
$Z_1Z_2\rightarrow \mathfrak{d}(7\vert \xi)$,
and $I_1I_2\rightarrow 1$, one can have the operator
\begin{eqnarray}
\hat{H}^{12\rightarrow3}&=&[\mathfrak{d}(7\vert\xi)+\gamma(\mathfrak{d}(5\vert\xi)
+\mathfrak{d}(6\vert\xi))]I_3   \nonumber\\
&+& [\mathfrak{d}(5\vert\xi)+\mathfrak{d}(6\vert\xi)+\gamma(\mathfrak{d}(7\vert\xi)+1)]Z_3\nonumber\\
&+&\delta[(\mathfrak{d}(1\vert\xi)-\mathfrak{d}(4\vert\xi))X_3-(\mathfrak{d}(2\vert\xi)+\mathfrak{d}(3\vert\xi))Y_3].\nonumber
\end{eqnarray}
From the constraint that $-1\leqslant \mathfrak{d}(j\vert\xi)\leqslant 1, j\in\{1,...,7\}$, it can be verified that
\begin{equation}
\vert\hat{H}^{12\rightarrow3}\vert^{\max}\leqslant 1+2\gamma+2\sqrt{(1+\gamma)^2+2\delta^2}.\nonumber
\end{equation}
Finally, because $\hat{H}$ is invariant under the relabelling the sites of the particles, one can obtain $\vert\hat{H}^{13\rightarrow 2}\vert^{\max}=\vert\hat{H}^{23\rightarrow 1}\vert^{\max}=\vert\hat{H}^{12\rightarrow 3}\vert^{\max}$ and $\vert\hat{H}^{2\rightarrow 13}\vert^{\max}=\vert\hat{H}^{3\rightarrow 12}\vert^{\max}=\vert\hat{H}^{1\rightarrow 23}\vert^{\max}$. According to the definition of $\beta_{\mathrm{GMST}}$ in
Eq. \eqref{betamax}, there is
\begin{equation}
\beta_{\mathrm{GMST}}=1+2\gamma+2\sqrt{(1+\gamma)^2+2\delta^2}.\nonumber
\end{equation}

\section*{Data Availability Statement}
Data sharing is not applicable to this article as no datasets were generated or analyzed during the current study.

\end{document}